\documentclass[conference]{IEEEtran}
\IEEEoverridecommandlockouts

\usepackage{cite}
\usepackage{amsmath,amssymb,amsfonts}
\usepackage{algorithmic}
\usepackage{graphicx}
\usepackage{textcomp}
\usepackage{xcolor}
\usepackage{tabularx}
\usepackage{booktabs}
\usepackage{multirow}
\usepackage{listings}
\usepackage{subfig}
\usepackage{balance}
\usepackage{xurl}
\usepackage{hyperref}

\hypersetup{
    colorlinks=true,
    linkcolor=black,
    filecolor=black,      
    urlcolor=black,
}

\def\BibTeX{{\rm B\kern-.05em{\sc i\kern-.025em b}\kern-.08em
    T\kern-.1667em\lower.7ex\hbox{E}\kern-.125emX}}

\newcommand{\hpcanalysis}{\texttt{hpcanalysis}}
\newcommand{\hpcreport}{\texttt{hpcreport}}
\newcommand{\hpcviewer}{\texttt{hpcviewer}}
\newcommand{\readapi}{\texttt{Read API}}
\newcommand{\queryapi}{\texttt{Query API}}
\newcommand{\dataanalysis}{\texttt{Data Analysis}}
\newcommand{\cpp}{\texttt{C++}}
\newcolumntype{Y}{>{\centering\arraybackslash}X}

\begin{document}

\title{Enhancing Performance Insight at Scale: A Heterogeneous Framework for Exascale Diagnostics}

\author{
\IEEEauthorblockN{Dragana Grbic}
\IEEEauthorblockA{
\textit{Department of Computer Science} \\
\textit{Rice University}\\
Houston, TX \\
dg76@rice.edu}
}

\maketitle

\begin{abstract}
As exascale systems reach unprecedented concurrency, traditional performance analysis tools struggle with the overhead of massive-scale telemetry. We present an accelerated infrastructure for the \hpcanalysis{} framework that leverages a high-performance \cpp{} API and GPU parallelism to enable high-throughput diagnostics. Our \cpp{} API achieves a 9.69-second ingestion time for 100,000 MPI ranks on Aurora. Furthermore, our GPU-accelerated layer achieves up to $314\times$ speedup over CPU-based processing when analyzing 100,000 execution traces. Finally, we implement a topology-aware workflow that maps logical performance outliers to physical Slingshot interconnect coordinates, localizing network congestion across 22 distinct racks on Aurora.

We also demonstrate how the framework's advanced interface seamlessly integrates with external tools to provide sophisticated analytical models. We introduce a novel tri-dimensional performance model that "re-materializes" iterative behavior from execution traces; using this model, we identified a 32.28\% potential speedup for a GAMESS workload on Frontier.
\end{abstract}

\begin{IEEEkeywords}
Exascale computing, Performance analysis, Heterogeneous frameworks, GPU acceleration, Interconnect congestion, Iteration-aware modeling.
\end{IEEEkeywords}

\section{Introduction}\label{sec:introduction}
Analyzing application performance on heterogeneous platforms remains a formidable challenge. While the High-Performance Computing (HPC) community has developed numerous tools for measuring and analyzing the performance of parallel scientific applications, many struggle to interpret the behavior of complex codes within critical scientific domains. These applications are particularly difficult to analyze because their interpretation requires orchestration across disparate parallel execution contexts. A primary hurdle lies in the sheer complexity of exascale platforms, where applications exhibit extreme levels of heterogeneity—spanning thousands of MPI tasks, intensive CPU multi-threading, and offloading heavy computations to GPUs.

Exascale supercomputers—such as Frontier~\cite{frontier_1, frontier_2} at ORNL, Aurora~\cite{aurora_1, aurora_2} at ANL, and El Capitan~\cite{elcapitan} at LLNL—provide cutting-edge infrastructures for executing diverse heterogeneous workloads. On such systems, users perform experiments across numerous parallel execution contexts, often distributed across thousands of compute nodes. The HPCToolkit~\cite{hpctoolkit_cpu, hpctoolkit_gpu} performance suite was developed to efficiently measure applications on exascale platforms, collecting fine-grained, instruction-level measurements of CPU and GPU code. While HPCToolkit provides the \hpcviewer{}~\cite{hpctoolkit_cpu} Graphical User Interface (GUI) for performance presentation, the recently introduced \hpcanalysis{} framework~\cite{hpcanalysis} was specifically designed to enable scalable, efficient, and interactive analysis of exascale measurements. Exascale measurements can reach massive scales, rendering manual GUI-based inspection time-consuming and often impractical.

In this paper, we focus on further enhancing the \hpcanalysis{} framework to support interactive, high-throughput analysis of large-scale applications. Specifically, we refine the initial architecture to increase data ingestion speed, leverage GPU-accelerated parallelism to offload the bulk of the computational analysis to specialized hardware, and integrate \hpcanalysis{} with external tools to enable advanced diagnostics. Finally, we demonstrate how this enhanced infrastructure efficiently interprets application performance at the full scale of modern supercomputers.

The paper makes the following contributions:
\begin{itemize}
    \item \textbf{Heterogeneous Infrastructure:} We present an accelerated \cpp{} and GPU-based substrate for the \hpcanalysis{} framework, enabling high-throughput ingestion and analysis of exascale performance measurements.
    \item \textbf{Tri-dimensional Modeling:} We introduce a novel performance model to "re-materialize" iterative behavior from execution traces, enabling the analysis of fine-grained imbalances and patterns typically lost in aggregate profiles.
    \item \textbf{Iteration-aware Diagnostics:} We demonstrate fine-grained analysis of GPU load imbalance in production workloads with repetitive behavior, providing mechanisms to predict significant performance gains.
    \item \textbf{Exascale Topology Mapping:} We evaluate our accelerated infrastructure on 100,000 MPI ranks on Aurora, localizing logical performance outliers to physical Slingshot interconnect coordinates to identify network congestion.
\end{itemize}

The remainder of the paper is organized as follows. Section~\ref{sec:related_work} reviews existing tools for performance measurement and analysis, providing background on challenges at exascale and the initial \hpcanalysis{} architecture. Section~\ref{sec:read_api} details the design of our high-throughput \cpp{} API for ingesting exascale measurements, while Section~\ref{sec:query_api} describes the GPU-accelerated abstraction layer we implemented to accelerate the analysis of extracted data. Section~\ref{sec:thicket} introduces the tri-dimensional performance model for iterative trace analysis, achieved by integrating \hpcanalysis{} with Thicket~\cite{thicket}. Section~\ref{sec:results} presents our experimental results on the Frontier and Aurora supercomputers. Finally, Section~\ref{sec:conclusion} concludes the paper.

\section{Background: \hpcanalysis{}}\label{sec:related_work}
The HPC community has developed numerious tools for measuring and analyzing the performance of parallel scientific applications. NVIDIA Nsight Systems~\cite{nsight_systems} enables system-wide tracing to identify communication and I/O bottlenecks, while NVIDIA Nsight Compute~\cite{nsight_compute} enables low-level analysis of individual GPU kernels. The TAU Performance System~\cite{tau} and Score-P~\cite{score_p} provide broad capabilities for instrumenting and collecting performance data across various programming models. Similarly, Caliper~\cite{caliper} and Extrae~\cite{extrae} provide mechanisms for intercepting runtime calls and integrating performance-aware annotations directly into source code. Scalasca~\cite{scalasca} and VampirTrace~\cite{vampir} collect trace data for visual analysis in specialized GUIs such as Paraver~\cite{paraver} and Vampir~\cite{vampir}.

Within this ecosystem, HPCToolkit stands out due to its powerful measurement methodology, which employs statistical sampling of hardware counters and call stack unwinding. This approach enables fine-grained, instruction-level measurement of CPU and GPU code with minimal execution overhead~\cite{hpctoolkit_refinement}. Such a sophisticated methodology facilitates the profiling and tracing of applications across thousands of compute nodes on heterogeneous exascale supercomputers—environments where traditional tools often struggle with the immense volume of measurement data.

Several tools have emerged to facilitate programmatic and automated post-mortem analysis of performance measurements. Hatchet~\cite{hatchet} is a Python library that can analyze performance profiles collected by various measurement tools. Hatchet models performance profiles as generic \texttt{GraphFrame} objects, which link a calling context tree to a Pandas~\cite{pandas} DataFrame that stores associated metrics. Thicket~\cite{thicket}, built upon Hatchet, extends these capabilities to support ensemble analysis across multiple experiments, enabling users to compare application performance across diverse platforms and scales. Pipit~\cite{pipit} is a Python library that can analyze execution traces from various file formats, such as Projections~\cite{projections} and Open Trace~\cite{open_trace}, by parsing them into consistent Pandas DataFrame structures.

However, as demonstrated in~\cite{hpcanalysis}, these tools exhibit critical inefficiencies when applied to fine-grained, large-scale HPCToolkit performance measurements. They load entire datasets into memory at once, which is infeasible for large-scale analysis. Furthermore, they don't provide techniques for pruning code regions with low information content or sampling subsets of massive performance profiles or execution traces. Notably, these tools store performance profiles in a dense format, which is highly inefficient for sparse profiles for GPU-accelerated applications; prior studies~\cite{hpctoolkit_ics} have shown that dense representations for these profiles can be over $1000\times$ larger than their sparse counterparts. Finally, these tools often analyze profiles and traces in isolation, preventing complex analysis tasks such reconstructing a calling context tree for a specific trace interval.

To bridge this gap, prior work~\cite{hpcanalysis} introduced the \hpcanalysis{} framework. This Python-based framework provides an advanced interface for the interactive, scalable, and efficient processing of fine-grained and large-scale performance measurements. The framework incorporates techniques for pruning code regions with low information content, techniques for sampling subsets of massive performance profiles and execution traces via specialized query expressions, and a unified approach to analyzing both profiles and traces. Previous studies~\cite{hpcanalysis} have shown that \hpcanalysis{} can efficiently process datasets from thousands of compute nodes by employing pruning and sampling.

\subsection{Architecture of the \hpcanalysis{} Framework}\label{sec:hpcanalysis}
Fig.~\ref{fig:hpcanalysis} illustrates the three-layered architecture of the \hpcanalysis{} framework as introduced in~\cite{hpcanalysis}. The framework consists of the \readapi{}, \queryapi{}, and \dataanalysis{} layers. This design decouples low-level data access from middle-level query expressions and high-level analytical tasks, optimizing each layer for maximum scalability and efficiency. The \readapi{} is a low-level API that parses and ingests requested data "slices" into memory. For small metadata, such as metric descriptions and profile descriptions, the \readapi{} parses the entire section (typically a few megabytes) upon the first request. For large metadata, such as the calling context tree, and large performance data, such as performance profiles execution traces, the \readapi{} employs pruning and sampling techniques:

\begin{figure*}
    \centering
    \includegraphics[width=\textwidth]{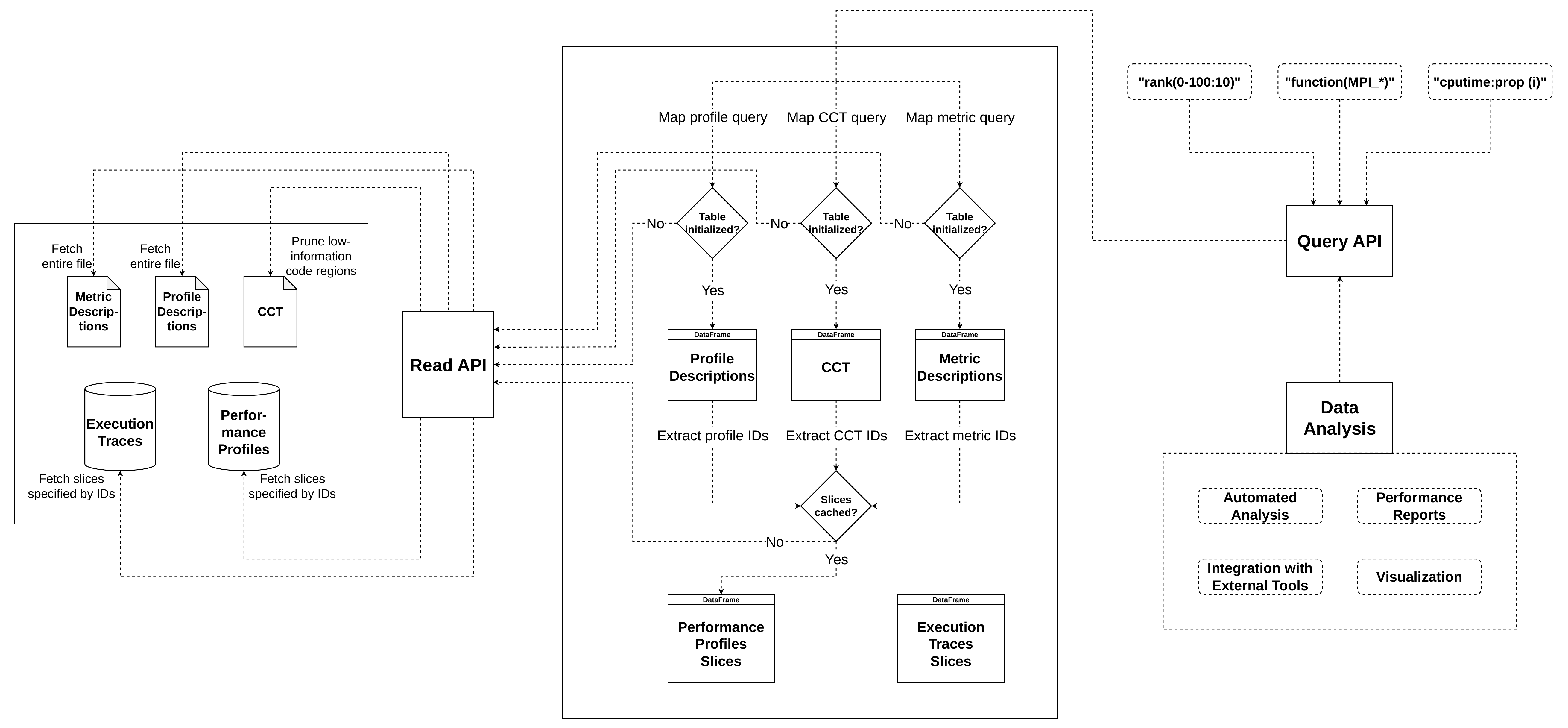}
    \caption{Architecture of the \hpcanalysis{} framework}
    \label{fig:hpcanalysis}
\end{figure*}

\begin{itemize}
    \item \textbf{Pruning Large Calling Context Trees:} HPCToolkit's global calling context tree represents the union of calling contexts across all parallel execution contexts. It contains fine-grained information, including library internals, line-level statements, and nodes with negligible inclusive costs. Because excessive detail can hamper analysis of critical bottlenecks, \hpcanalysis{} enables users to declare strategies for pruning code regions with low information content from the global calling context tree. Pruning strategies are evaluated once on the summary profile - an aggregate of all parallel performance profiles - and then propagated across parallel profiles. The \readapi{} leverages the internal structure of performance files, where calling context tree nodes and metrics are sorted by their IDs, to extract only requested "slices" via selective access/binary search, without exploring massive profiles in their entirety. Furthermore, the \readapi{} employs internal parallelism to parse individual profiles simultaneously, significantly accelerating data ingestion.
    \item \textbf{Sampling Massive Profiles and Traces:} To facilitate analysis of large-scale datasets, the \readapi{} enables users to sample subsets of massive performance profiles and execution traces to estimate results on smaller subsets. Performance profiles consist of local calling context trees annotated with metric values for their parallel execution context. Given their vast volume at scale, the \readapi{} does not fetch the entire section upon first access; instead, it uses indices provided by the \queryapi{} to extract "slices". Users provides queries specifying execution contexts (ranges of MPI ranks/threads), calling contexts (specific call chains or functions), and metrics of interest, which are mapped to raw indices for the \readapi{}. Within the database, profiles are organized as an array indexed by profile IDs, with internal calling context tree nodes and metrics sorted by their respective IDs. This enables the \readapi{} to utilize indexing, binary search, and parallelism to perform sampling. Execution traces, representing series of timed calling context tree node events, are handled similarly.
\end{itemize}

The \queryapi{} is a middle-level API that enables users to access performance data via specialized query expressions rather than the raw indices required by the \readapi{}. It orchestrates out-of-core data handling through two primary mechanisms:

\begin{itemize}
    \item \textbf{Query Mapping:} Users submit queries specifying "slices" of interest. For instance, when analyzing an execution on 100,000 MPI ranks, users can submit queries \texttt{"rank(0-100000:100)"}, \texttt{"function(MPI\_*)"}, and \texttt{"cputime:prop (i)"} to examine CPU performance of MPI routines for every 100th rank. These queries are mapped to raw indices using metadata tables and then processed by the \readapi{} using parallelism, indexing, and binary search. Similarly, users can provide rank queries and time intervals to perform sampling of execution traces.
    \item \textbf{Caching and Requesting:} Extracted data "slices" are stored in Pandas DataFrame tables for analysis. When data is requested, the \queryapi{} checks if the corresponding logical IDs are already present/cached in its internal DataFrame tables. If not, it requests the missing "slices" from the \readapi{}. Once loaded, users can leverage standard DataFrame operations to analyze program behavior and performance.
\end{itemize}

The \dataanalysis{} layer, built atop the \queryapi{}, initially provided utilities for generating flat profiles, detecting load imbalances, and visualizing calling context trees, alongside the specialized \hpcreport{} utility~\cite{hpcanalysis}. \hpcreport{} analyzes the summary profile to categorize the execution time into domains (e.g., MPI or GPU execution) and automatically infers whether an application is compute, memory, communication, or I/O bound.

In the following sections, we describe our enhancements to the initial \hpcanalysis{}' infrastructure to enable more scalable and efficient processing of exascale measurements and introduce novel workflows for advanced, automated performance interpretation at scale.

\section{Accelerating \readapi{} for Performance Data Extraction}\label{sec:read_api}
The initial architecture presented in Fig.~\ref{fig:hpcanalysis} was implemented entirely in Python, including \readapi{}, \queryapi{}, and \dataanalysis{} layers. These APIs were designed to integrate seamlessly into Jupyter~\cite{jupyter} notebooks, enabling users to write queries and specialized analysis tasks to investigate application performance. When processing massive exascale measurements, the original implementation relied on techniques for pruning code regions with low information content and sampling subsets of massive performance profiles and execution traces via specialized query expressions. While they effectively reduce the volume of exascale measurements by sacrificing some level of detail, researchers often require full or larger datasets to perform high-fidelity diagnostics with greater confidence.

Techniques for pruning calling context trees described in Section~\ref{sec:hpcanalysis} are highly efficient, as they identify major bottlenecks on the summary profile and propagate them across massive parallel performance profiles via selective access/binary search. However, sampling subsets of massive performance profiles and execution traces can sometimes omit critical performance information, preventing a comprehensive examination of the execution. While prior work~\cite{hpcanalysis} successfully identified communication anomalies in a 64K-rank LAMMPS~\cite{lammps} execution on Frontier by sampling a small subset of ranks, this approach was largely motivated by the ingestion bottlenecks inherent in the Python-based \readapi{}. To enable users to analyze their large-scale executions with a greater level of detail, we focus on providing an accelerated infrastructure for ingesting exascale measurements.

Because data ingestion represented a bottleneck, our first enhancement to \hpcanalysis{} was to re-implement the \readapi{} as a shared \cpp{} library. This new API follows the exact interface described in~\cite{hpcanalysis}, ensuring backward compatibility while significantly improving throughput. The initial Python \readapi{} relied on standard file operations (\texttt{file.seek()} and \texttt{file.read()}), which typically involve double-copying data: first from the disk to the kernel buffer, and then from the kernel buffer to the Python process memory space. In our new \cpp{} API, we utilize the \texttt{mmap}~\cite{mmap} system call to map files directly into the process's virtual address space. This enables the operating system to load data lazily upon access; subsequent accesses use simple pointer arithmetic, which is significantly faster than repeated file I/O calls. While Python provides an \texttt{mmap} module, it is notably more efficient in \cpp{}, where address jumps are performed using single machine instructions without the overhead of object creation or interpreter intervention.

Furthermore, the \cpp{} API leverages native OpenMP parallelism, which is effectively unavailable in Python due to the Global Interpreter Lock (GIL). The GIL is a mutex that prevents multiple native threads from executing Python bytecodes simultaneously. While \texttt{joblib.Parallel} used in the initial Python implementation was useful for I/O-bound tasks, where the GIL can be released during disk operations, for CPU-bound tasks \texttt{joblib}~\cite{joblib} often falls back to multiprocessing, incurring the overhead of process creation and inter-process communication. In contrast, our \cpp{} API uses fast, shared-memory OpenMP threads for low-overhead parallel processing of multiple performance profiles and execution traces. By default, the \cpp{} API utilizes maximum hardware concurrency to determine the number of OpenMP threads, while the user retains the flexibility to specify a custom value. Finally, we use the \texttt{pybind11}~\cite{pybind11} library to convert native \cpp{} objects to Python-compatible structures, ensuring that the bulk of the computational work is completed in \cpp{} before any conversion overhead is incurred. In Section~\ref{sec:amg}, we evaluate our \cpp{} API on a massive-scale execution on Aurora, demonstrating that it can ingest exascale measurements with significantly higher throughput than the original Python-based \readapi{}.

\section{Accelerating \queryapi{} for Performance Data Analysis}\label{sec:query_api}

\begin{table*}
\centering
\caption{Performance Comparison of Pandas (CPU) and cuDF (NVIDIA GPU) DataFrame Operations on Polaris}
\label{tab:cudf_performance}
\renewcommand{\arraystretch}{1.5}
\resizebox{\textwidth}{!}{%
\begin{tabular}{l | ccc | ccc | ccc | ccc}
    \noalign{\hrule height 1pt}
    \textbf{Number of} & \multicolumn{3}{c|}{\textbf{Grouping (s)}} & \multicolumn{3}{c|}{\textbf{Sorting (s)}} & \multicolumn{3}{c|}{\textbf{Filtering (s)}} & \multicolumn{3}{c}{\textbf{Merging (s)}} \\
    \textbf{Traces} & CPU & GPU & Speedup & CPU & GPU & Speedup & CPU & GPU & Speedup & CPU & GPU & Speedup \\
    \hline
        
    \textbf{10} & 0.001667 & 0.000982 & 1.70$\times$ & 0.000233 & 0.000581 & 0.40$\times$ & 0.000372 & 0.003548 & 0.10$\times$ & 0.006814 & 0.001739 & 3.92$\times$ \\
    \noalign{\smallskip}
    
    \textbf{100} & 0.006041 & 0.001980 & 3.05$\times$ & 0.001751 & 0.000520 & 3.37$\times$ & 0.000508 & 0.001278 & 0.40$\times$ & 0.008197 & 0.002199 & 3.73$\times$ \\
    \noalign{\smallskip}
    
    \textbf{1,000} & 0.034233 & 0.002260 & 15.15$\times$ & 0.024794 & 0.002224 & 11.15$\times$ & 0.002622 & 0.002433 & 1.08$\times$ & 0.028052 & 0.005326 & 5.27$\times$ \\
    \noalign{\smallskip}
    
    \textbf{10,000} & 0.483654 & 0.003993 & 121.13$\times$ & 0.352657 & 0.003165 & 111.44$\times$ & 0.033414 & 0.003908 & 8.55$\times$ & 0.318041 & 0.007574 & 41.99$\times$ \\
    \noalign{\smallskip}
    
    \textbf{100,000} & 7.696614 & 0.024470 & 314.54$\times$ & 4.856504 & 0.023193 & 209.40$\times$ & 0.375433 & 0.008901 & 42.18$\times$ & 3.125140 & 0.046568 & 67.11$\times$ \\
    
    \noalign{\hrule height 1pt}
\end{tabular}%
}
\end{table*}

\begin{table*}
\centering
\caption{Performance Comparison of Pandas (CPU) and hipDF (AMD GPU) DataFrame Operations on Frontier}
\label{tab:hipdf_performance}
\renewcommand{\arraystretch}{1.5}
\resizebox{\textwidth}{!}{%
\begin{tabular}{l | ccc | ccc | ccc | ccc}
    \noalign{\hrule height 1pt}
    \textbf{Number of} & \multicolumn{3}{c|}{\textbf{Grouping (s)}} & \multicolumn{3}{c|}{\textbf{Sorting (s)}} & \multicolumn{3}{c|}{\textbf{Filtering (s)}} & \multicolumn{3}{c}{\textbf{Merging (s)}} \\
    \textbf{Traces} & CPU & GPU & Speedup & CPU & GPU & Speedup & CPU & GPU & Speedup & CPU & GPU & Speedup \\
    \hline
        
    \textbf{10} & 0.000832 & 0.001015 & 0.82$\times$ & 0.000207 & 0.000453 & 0.46$\times$ & 0.000339 & 0.001405 & 0.24$\times$ & 0.006900 & 0.001523 & 4.53$\times$ \\
    \noalign{\smallskip}
    
    \textbf{100} & 0.003507 & 0.000976 & 3.59$\times$ & 0.001751 & 0.000484 & 3.62$\times$ & 0.000579 & 0.001384 & 0.42$\times$ & 0.008702 & 0.001562 & 5.57$\times$ \\
    \noalign{\smallskip}
    
    \textbf{1,000} & 0.036337 & 0.001744 & 20.84$\times$ & 0.026613 & 0.000943 & 28.22$\times$ & 0.003052 & 0.001680 & 1.82$\times$ & 0.031599 & 0.002735 & 11.55$\times$ \\
    \noalign{\smallskip}
    
    \textbf{10,000} & 0.575268 & 0.007235 & 79.51$\times$ & 0.439099 & 0.003317 & 132.39$\times$ & 0.056606 & 0.002421 & 23.38$\times$ & 0.448289 & 0.008239 & 54.41$\times$ \\
    \noalign{\smallskip}
    
    \textbf{100,000} & 9.342253 & 0.060572 & 154.23$\times$ & 6.452984 & 0.029318 & 220.10$\times$ & 0.814406 & 0.006581 & 123.75$\times$ & 5.459309 & 0.065223 & 83.70$\times$ \\
    
    \noalign{\hrule height 1pt}
\end{tabular}%
}
\end{table*}

\begin{table*}
\centering
\caption{Performance Comparison of Pandas (CPU) and SYCL (Intel GPU) Array Operations on Aurora}
\label{tab:sycl_performance}
\renewcommand{\arraystretch}{1.5}
\resizebox{\textwidth}{!}{%
\begin{tabular}{l | ccc | ccc | ccc | ccc | ccc}
    \noalign{\hrule height 1pt}
    \textbf{Operation} & \multicolumn{3}{c|}{\textbf{10 Traces}} & \multicolumn{3}{c|}{\textbf{100 Traces}} & \multicolumn{3}{c|}{\textbf{1,000 Traces}} & \multicolumn{3}{c|}{\textbf{10,000 Traces}} & \multicolumn{3}{c}{\textbf{100,000 Traces}} \\
    \textbf{(s)} & CPU & GPU & Speedup & CPU & GPU & Speedup & CPU & GPU & Speedup & CPU & GPU & Speedup & CPU & GPU & Speedup \\
    \hline
    
    \textbf{Scalar Compare} & 0.000043 & 0.000122 & 0.35$\times$ & 0.000071 & 0.000163 & 0.44$\times$ & 0.000269 & 0.000134 & 2.01$\times$ & 0.001706 & 0.000206 & 8.28$\times$ & 0.013758 & 0.001196 & 11.50$\times$ \\
    \noalign{\smallskip}
    
    \textbf{Vector Add} & 0.000062 & 0.000146 & 0.42$\times$ & 0.000073 & 0.000142 & 0.51$\times$ & 0.000624 & 0.000131 & 4.76$\times$ & 0.000882 & 0.000590 & 1.49$\times$ & 0.010708 & 0.001199 & 8.93$\times$ \\
    \noalign{\smallskip}
    
    \textbf{In-place Multi} & 0.000075 & 0.000123 & 0.61$\times$ & 0.000094 & 0.000121 & 0.78$\times$ & 0.000790 & 0.000114 & 6.93$\times$ & 0.004231 & 0.000167 & 25.34$\times$ & 0.035281 & 0.000559 & 63.11$\times$ \\
    \noalign{\smallskip}
    
    \textbf{Sorting} & 0.000036 & 0.000133 & 0.27$\times$ & 0.000322 & 0.000151 & 2.13$\times$ & 0.003722 & 0.000292 & 12.75$\times$ & 0.034060 & 0.001461 & 23.31$\times$ & 0.485844 & 0.020705 & 23.47$\times$ \\
    \noalign{\smallskip}

    \textbf{Filtering} & 0.000013 & 0.000287 & 0.05$\times$ & 0.000054 & 0.000289 & 0.19$\times$ & 0.000441 & 0.000253 & 1.74$\times$ & 0.003502 & 0.000712 & 4.92$\times$ & 0.125687 & 0.002120 & 59.29$\times$ \\
    \noalign{\smallskip}

    \textbf{Reduction Sum} & 0.000009 & 0.000076 & 0.12$\times$ & 0.000015 & 0.000057 & 0.26$\times$ & 0.000109 & 0.000058 & 1.88$\times$ & 0.000872 & 0.000070 & 12.46$\times$ & 0.022190 & 0.000324 & 68.49$\times$ \\
    \noalign{\smallskip}
    
    \textbf{Cumulative Sum} & 0.000017 & 0.000097 & 0.18$\times$ & 0.000105 & 0.000078 & 1.35$\times$ & 0.000864 & 0.000088 & 9.82$\times$ & 0.007594 & 0.000521 & 14.58$\times$ & 0.120441 & 0.001474 & 81.71$\times$ \\
        
    \noalign{\hrule height 1pt}
\end{tabular}%
}
\end{table*}

After enhancing the low-level \readapi{}, we turned our focus to the middle-level \queryapi{}, which enables users to access performance data via specialized query expressions. In \hpcanalysis{}' initial implementation, \queryapi{} stored extracted performance data in Pandas DataFrame tables for further analysis. While Pandas leverages efficient, vectorized operations through \texttt{numpy}~\cite{numpy}, it is restricted to a single-threaded CPU execution, hindering scalable processing of exascale measurements. To address this, we accelerated the analysis workflow by replacing Pandas DataFrame tables with a GPU-accelerated abstraction layer, enabling high-throughput parallel processing across heterogeneous architectures.

We implemented the \texttt{HpcDataFrame} abstraction, a unified interface that the \queryapi{} utilizes to store and manipulate extracted performance data. By default, the \texttt{HpcDataFrame} executes on the CPU using a Pandas backend. However, if the user explicitly enables GPU acceleration, the framework dynamically selects a specialized backend based on the detected hardware. On systems with NVIDIA GPUs, the framework utilizes the \texttt{cuDF}~\cite{cudf} library, which provides a GPU-accelerated, Pandas-compatible interface. Similarly, on systems with AMD GPUs, the framework utilizes the \texttt{hipDF}~\cite{hipdf} library. Because a native, Pandas-compatible library for Intel GPUs—comparable to \texttt{cuDF} or \texttt{hipDF}—is not available, we implemented our custom SYCL-based wrapper to provide GPU-accelerated support for Intel architectures. Utilizing the \texttt{dpnp}~\cite{dpnp} library, we modeled the \texttt{HpcDataFrame} by storing each table column as a discrete array in Intel GPU memory. This wrapper implements the core functionalities required by the \queryapi{} by leveraging SYCL-accelerated kernels. While the SYCL wrapper currently focuses on single-column operations, it provides the foundational infrastructure needed for high-throughput analysis on Intel GPUs. For complex multi-column operations that exceed the current SYCL wrapper's capabilities, users retain the flexibility to cast the \texttt{HpcDataFrame} back to a CPU-based Pandas representation for final processing. \texttt{HpcDataFrame} abstraction enables the \queryapi{} to interact with a single, consistent API, while the abstraction itself manages hardware-specific execution.

We evaluated the performance of the \texttt{HpcDataFrame} and its GPU-accelerated backends by analyzing execution traces from a large-scale execution of the AMG benchmark~\cite{amg} on Aurora. We performed the execution across 1,000 compute nodes using 100 ranks per node for a total of 100,000 MPI ranks. The benchmark solved a large-scale Laplace problem using a local grid size of $40 \times 50 \times 50$ per rank, which generated a performance database of approximately 11~GB. We analyzed the processing of execution traces across three distinct platforms, leveraging vendor-specific software stacks for GPU acceleration: Polaris~\cite{polaris_1, polaris_2} (NVIDIA A100 40GB GPUs) using CUDA 12.8~\cite{cuda}; Frontier (AMD Instinct MI250X GPUs) using ROCm 7.0.2~\cite{rocm}; and Aurora (Intel Max 1550 GPUs) using the Intel oneAPI~\cite{oneapi} environment operating over the Level Zero~\cite{levelzero} 12.60.7 driver interface. On each platform, we measured the time required to manipulate trace events using the standard CPU-based Pandas backend against the corresponding GPU-accelerated backend (\texttt{cuDF} for Polaris, \texttt{hipDF} for Frontier, and our SYCL wrapper for Aurora). We evaluated manipulating data subsets ranging from 10 to 100,000 MPI ranks. To ensure our results account for potential noise in tested environments, every operation reported in Tables~\ref{tab:cudf_performance}, \ref{tab:hipdf_performance}, and \ref{tab:sycl_performance} was executed 10 times, with the presented metrics representing the calculated average across those runs.

The experimental results demonstrate a significant performance advantage when utilizing GPU-accelerated backends for large-scale analysis. Across all platforms, we observe a consistent trend: for smaller datasets (e.g., 10 to 100 traces), the CPU often outperforms the GPU due to the overhead associated with memory transfers and kernel launches. However, as the data volume scales toward 100,000 traces, the massive parallelism of the GPU becomes transformative. The most remarkable speedups were achieved with \texttt{cuDF} on Polaris and \texttt{hipDF} on Frontier; for high-complexity operations like grouping and sorting, we observed speedups exceeding 314$\times$ and 220$\times$, respectively. For the Intel-based Aurora system (Table~\ref{tab:sycl_performance}), we evaluated speedups on the fundamental array operations that underpin our SYCL wrapper. Operations such as cumulative sum and reduction sum saw speedups of up to 81$\times$ and 68$\times$, respectively. These findings validate the impact of our \texttt{HpcDataFrame} abstraction and serve as a powerful benchmark for GPU-accelerated libraries on real-world exascale performance data.

\section{Trace Analysis With Thicket}\label{sec:thicket}

\begin{figure*}
  \centering
  \includegraphics[width=\textwidth]{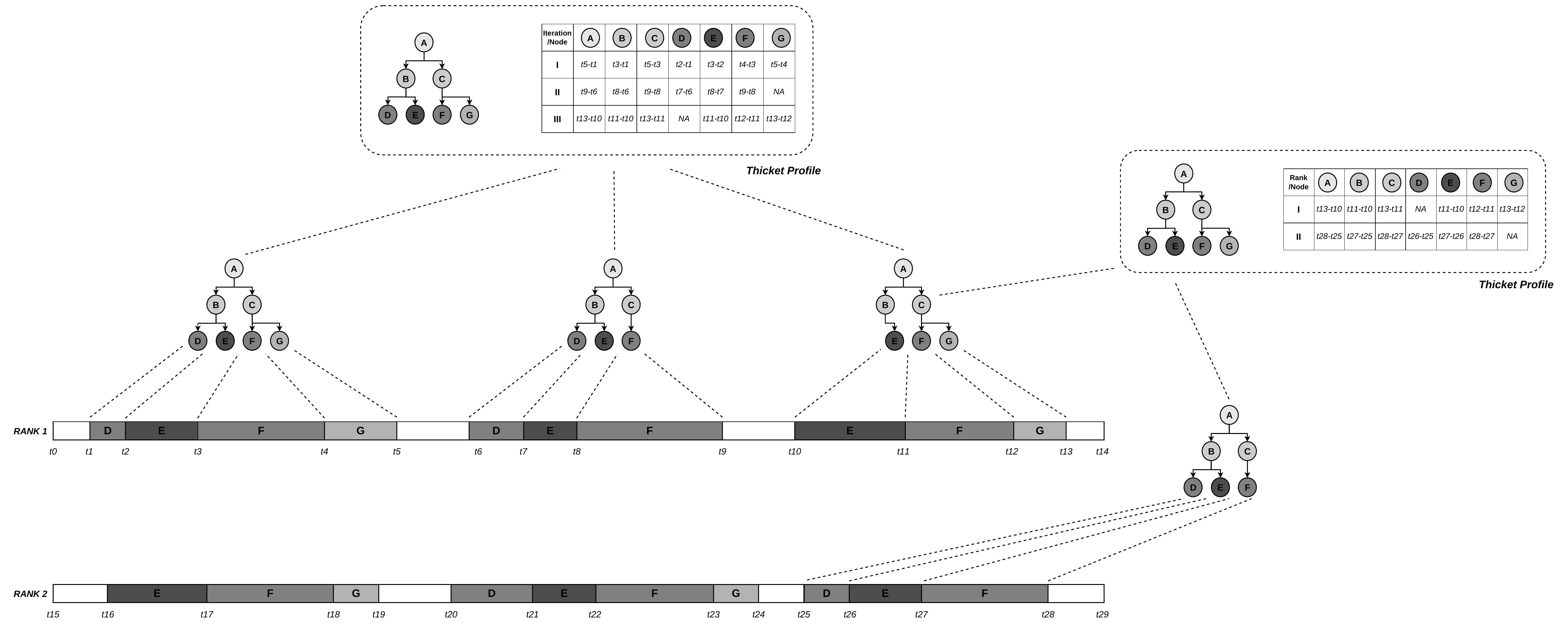}
  \caption{Workflow for iteration-aware trace analysis. Traces are scanned to detect iterative boundaries, followed by reconstructing calling context trees for each interval. The resulting data is merged into a three-dimensional Thicket profile (nodes $\times$ traces $\times$ iterations) for fine-grained comparative analysis.}
  \label{fig:thicket}
\end{figure*}

In Section~\ref{sec:related_work}, we noted that Hatchet~\cite{hatchet} and Thicket~\cite{thicket} tools provide generic models for storing and analyzing calling context trees and performance profiles collected by various measurement tools. While Hatchet analyzes performance profiles from a single application run, Thicket provides a multi-dimensional model that can compare performance profiles across numerous experiments. Thicket utilizes a two-dimensional infrastructure: it maintains a unified calling context tree representing the union of calling contexts across all experiments and maps the union tree to a DataFrame where each row corresponds to a specific experiment. However, as discussed in Section~\ref{sec:related_work}, Hatchet and Thicket scale poorly for fine-grained, large-scale performance data, lack support for pruning and sampling exascale measurements, and do not provide native support for analyzing execution traces.

In this work, we integrate the advanced \hpcanalysis{}' interface with Thicket’s multi-dimensional analytical model to provide workflows for advanced trace analysis. Our goal is to evolve Thicket’s initial two-dimensional model for analyzing performance profiles across experiments into a three-dimensional model for analyzing fine-grained behaviors within and across execution traces, such as iterative patterns and load imbalances. By leveraging \hpcanalysis{}' interface to efficiently extract critical performance "slices", we can populate a Thicket model that enables users to compare execution phases within and across execution traces using a few lines of Python code. This integration demonstrates how the \hpcanalysis{} infrastructure can extend external tools to provide sophisticated models for advanced performance analysis.

We implemented an advanced workflow, shown in Fig.~\ref{fig:thicket}, that detects iterative phases within and across execution traces to construct a unified Thicket model. For a given set of execution traces, the workflow first identifies sequences of trace events that represent repeated iterative phases. Once they are identified, \hpcanalysis{} reconstructs a distinct Hatchet profile for each individual iteration. This step is crucial because HPCToolkit’s default performance profiles collapse the time dimension, aggregating the total cost of each calling context across the entire execution. To perform fine-grained iterative analysis, one must "re-materialize" performance profiles for specific time intervals (iterations) from the execution traces—a core capability of our new workflow.

After reconstructing Hatchet profiles for individual iterations, we aggregate them into a unified Thicket object. Unlike standard Thicket object designed to compare distinct experiments, our workflow constructs a tri-dimensional model that facilitates comparisons within and across execution traces. We achieve this by implementing a tri-dimensional index within Thicket’s DataFrame table, consisting of: (1) the event (calling context tree node), (2) the trace index, and (3) the iteration index. The DataFrame columns store the actual time spent in each event for the specific trace and iteration. Using standard DataFrame operations, users can slice this model by fixing a specific iteration to compare it across execution traces, or by fixing a specific trace to analyze performance evolution across its iterations. This enables the detection of load imbalances and the prediction of potential performance gains, which we discuss in Section~\ref{sec:gamess}.

\section{Case Studies}\label{sec:results}
In this section, we present two case studies demonstrating how the enhanced \hpcanalysis{} infrastructure enables advanced analysis and interpretation of program performance at scale. First, we evaluate our novel performance model for analyzing iterative phases within and across execution traces using a real-world application. Second, we demonstrate how the framework's accelerated infrastructure can process massive-scale executions to discover critical performance inefficiencies. The first experiment was performed on the Frontier supercomputer (AMD EPYC 7A53 CPUs, AMD Instinct MI250X GPUs, HPE Slingshot 11) and the second on the Aurora supercomputer (Intel Xeon Max 9470 CPUs, Intel Max 1550 GPUs, HPE Slingshot 11).

\subsection{GAMESS GPU Utilization on Frontier}\label{sec:gamess}
In our first case study, we analyzed a single-node execution of a Hartree-Fock energy calculation in GAMESS~\cite{gamess} on Frontier. GAMESS is a quantum chemistry suite used to calculate the electronic structure of nanoparticles. We performed the execution across 16 MPI ranks; in a typical GAMESS configuration, each computational rank offloads to one GPU and is paired with a data-server rank. Consequently, the total number of ranks is double the number of GPUs. We utilized this execution to evaluate our tri-dimensional model's ability to examine iterative phases within and across execution traces and quantify GPU utilization in detail.

\begin{figure}
    \centering
    \includegraphics[width=.5\textwidth]{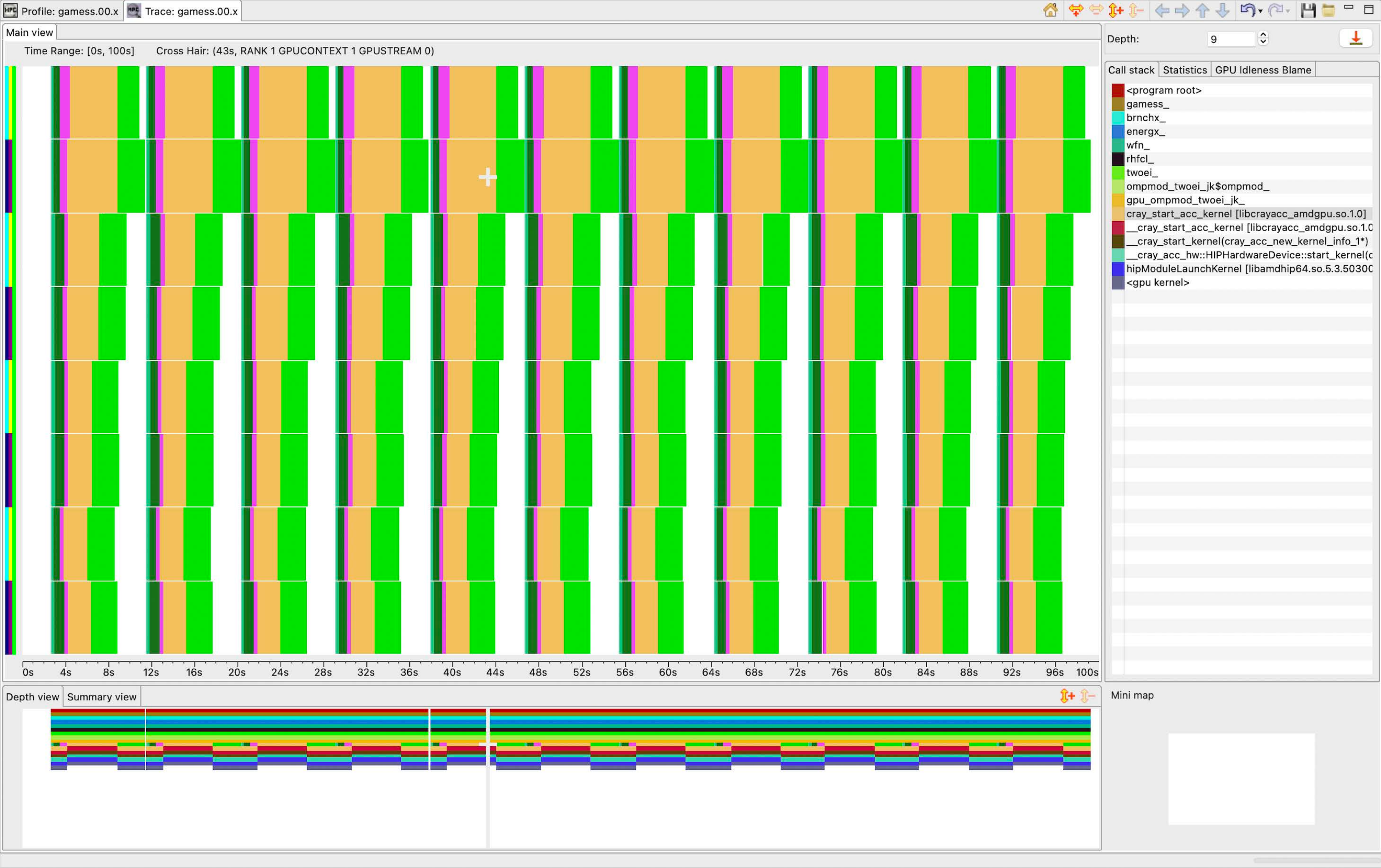}
    \caption{Single-node GAMESS execution on Frontier. The varying GPU stream lengths and idle gaps illustrate significant load imbalance across MPI ranks.}
    \label{fig:gamess_traces}
\end{figure}

First, in Fig.~\ref{fig:gamess_traces}, we visualized the GPU execution traces for the eight computational MPI ranks using \hpcviewer{}. Visual inspection revealed a significant load imbalance, with several GPUs spending considerable time idle. This stems from a triangularly nested loop, where computational tasks are assigned varying amounts of work. While the GUI highlights the gaps, it cannot quantify load balance ratios for specific GPU kernels or predict potential performance gains. We used this execution to demonstrate how \hpcanalysis{} and our tri-dimensional model can automatically detect load imbalances and convert them into actionable insights.

We implemented a workflow that performs fine-grained analysis of GPU utilization for the GAMESS execution on Frontier. The process follows several key steps:

\begin{itemize}
    \item \textbf{GPU Metrics Identification:} \hpcanalysis{} detects which GPU metrics (e.g., kernel execution, data copies, memory allocation, or synchronization) contain non-zero samples. For the GAMESS execution on Frontier, it identified GPU kernel execution (\texttt{gker}) and explicit data copies (\texttt{gxcopy}) as the primary metrics of interest.
    \item \textbf{GPU Kernels Identification:} Because GPU performance profiles are sparse, \hpcanalysis{} extracts GPU kernels by sampling the summary profile for calling contexts attributed with the identified GPU metrics. For the GAMESS execution on Frontier, it isolated six primary GPU kernels (e.g., \texttt{gpu\_rhf\_j05\_ppps\_}) and their parent \texttt{gpu\_ompmod\_twoei\_jk\_}.
    \item \textbf{GPU Imbalance Quantification:} \hpcanalysis{} examines parallel performance profiles to calculate a global balance ratio for each GPU kernel, defined as the average execution time across ranks divided by the maximum. As shown in Table~\ref{tab:gamess_imbalance}, several GPU kernels exhibited severe imbalance, with ratios as low as 0.50 for \texttt{gpu\_rhf\_j04\_psps\_} (where a ratio of 1.0 represents perfect balance).
    \item \textbf{Iterative Trace Analysis with Thicket:} Because performance profiles aggregate costs, execution traces are required to understand variance over time. \hpcanalysis{} scans GPU execution traces to detect iterative boundaries, reconstructs separate Hatchet profiles for each iteration, and merges them into a tri-dimensional Thicket model described in Section~\ref{sec:thicket}. This enables the framework to orchestrate a complex analysis: first using the summary profile to detect major bottlenecks, then using parallel performance profiles to quantify global load imbalance, and finally using execution traces to understand behavior across iterations.
\end{itemize}

\begin{table*}[t]
\centering
\caption{Analysis of global load imbalance and iteration variance for the major GPU kernels in GAMESS execution on Frontier.}
\label{tab:gamess_imbalance}
\begin{tabularx}{\textwidth}{@{} Y Y Y Y Y @{}}
\toprule
\textbf{GPU Kernel} & \textbf{Execution} & \textbf{Global} & \textbf{Across-Rank} & \textbf{Within-Rank} \\
& \textbf{Share (\%)} & \textbf{Balance Ratio} & \textbf{Time CV (\%)} & \textbf{Time CV (\%)} \\
\midrule
gpu\_rhf\_j05\_ppps\_    & 42.89\% & 0.64 & 34.21 & 1.45 \\ 
gpu\_rhf\_j06\_pppp\_    & 35.93\% & 0.96 &  7.20 & 1.36 \\ 
gpu\_rhf\_j03\_ppss\_    & 10.63\% & 0.79 & 19.84 & 7.79 \\ 
gpu\_rhf\_j04\_psps\_    & 6.99\%  & 0.50 & 46.31 & 2.66 \\ 
gpu\_rhf\_j02\_psss\_    & 3.46\%  & 0.80 & 19.68 & 9.09 \\ 
gpu\_rhf\_j01\_ssss\_    & 0.10\%  & 0.68 & 27.51 & 6.28 \\ 
\bottomrule
\end{tabularx}
\end{table*}

\begin{table*}[t]
\centering
\caption{Estimated wall-clock time reductions through ideal load balancing of GPU kernels in GAMESS execution on Frontier.}
\label{tab:gamess_savings}
\begin{tabularx}{\textwidth}{@{} Y Y Y Y Y @{}}
\toprule
\textbf{GPU Kernel} & \textbf{Avg. Mean} & \textbf{Avg. Max} & \textbf{Potential Savings} & \textbf{Total Estimated} \\
& \textbf{Time (s)} & \textbf{Time (s)} & \textbf{(s / Iteration)} & \textbf{Reduction (s)} \\
\midrule
gpu\_rhf\_j05\_ppps\_ & 2.963 & 4.650 & 1.687 & 18.557 \\ 
gpu\_rhf\_j06\_pppp\_ & 2.483 & 2.599 & 0.116 &  1.276 \\ 
gpu\_rhf\_j03\_ppss\_ & 0.734 & 0.945 & 0.211 &  2.321 \\ 
gpu\_rhf\_j04\_psps\_ & 0.483 & 0.958 & 0.475 &  5.225 \\ 
gpu\_rhf\_j02\_psss\_ & 0.239 & 0.300 & 0.061 &  0.671 \\ 
gpu\_rhf\_j01\_ssss\_ & 0.007 & 0.010 & 0.003 &  0.033 \\ 
\bottomrule
\end{tabularx}
\end{table*}

\begin{figure}
    \centering
    \includegraphics[width=.35\textwidth]{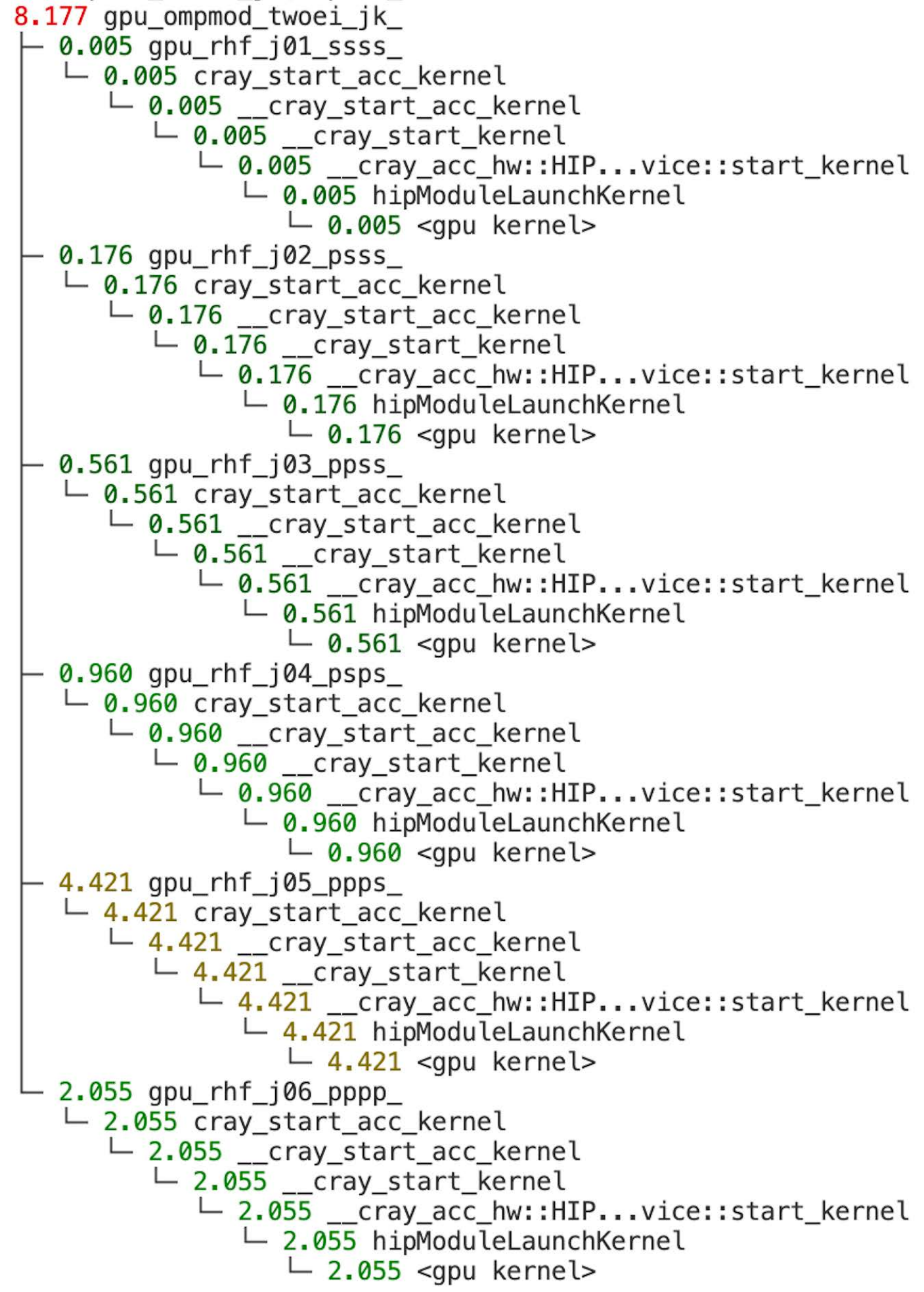}
    \caption{Thicket model for a single iteration (rank 0, iteration 0) for the GAMESS execution on Frontier.}
    \label{fig:gamess_thicket}
\end{figure}

Fig.~\ref{fig:gamess_thicket} illustrates a single iteration (rank 0, iteration 0) within the Thicket model for the GAMESS execution on Frontier. While Thicket stores data for all ranks and iterations in its tri-dimensional DataFrame index, the visualization enables users to select specific "slices". This structure enables powerful, one-line statistical queries. For example:

\begin{lstlisting}[language=Python]
df_kernel.groupby("iteration")["time"].agg(
    ["mean", "min", "max", "std"]
)
\end{lstlisting}

This query calculates load imbalance across MPI ranks for every iteration for a specific GPU kernel. By swapping \texttt{"iteration"} for \texttt{"rank"}, users can examine variance within a rank over time. Using these queries, we calculated the Coefficient of Variation (CV) for major GPU kernels, as shown in Table~\ref{tab:gamess_imbalance}. We found that while CV across ranks reached as high as 46.31\%, the CV within ranks remained under 9\%, proving that the iterative behavior is stable but poorly distributed across resources.

Finally, we used the tri-dimensional model to predict potential performance gains under ideal load balancing. By calculating the difference between the average maximum time and the average mean time per iteration for each GPU kernel (Table~\ref{tab:gamess_savings}), we estimated potential savings of 28.08 seconds across 11 iterations. Given the total application time of 87 seconds, resolving this imbalance could yield an approximately 32.28\% speedup. This demonstrates how \hpcanalysis{} can transform raw execution traces into concrete optimization targets. Our tri-dimensional model targets applications with regular, iterative patterns to analyze load imbalance within and across execution traces; it is not optimized for complex applications with irregular or highly variable execution phases.

\subsection{Exascale Network Congestion Analysis on Aurora}\label{sec:amg}
In our second case study, we demonstrate how the accelerated \hpcanalysis{} infrastructure enables efficient processing of massive-scale executions to isolate critical performance inefficiencies. We analyzed the 100,000-rank execution of the AMG benchmark on Aurora, which served as a stress test for our \texttt{HpcDataFrame} abstraction in Section~\ref{sec:query_api}. This 100,000-rank execution was distributed across 1,000 compute nodes, with 100 MPI ranks per node. Here, we evaluate how our accelerated infrastructure can efficiently process exascale measurements to localize network-level bottlenecks.

\textbf{High-Throughput Ingestion of Exascale Data.} The AMG execution on Aurora produced 100,000 parallel profiles and 100,000 execution traces. The global calling context tree—the union of local calling context trees across all MPI ranks—contained 97,833 distinct calling contexts. Consequently, the resulting performance dataset represents a massive, structured matrix of up to $100,000 \times 97,833$ calling contexts, each annotated with multiple inclusive and exclusive costs. After applying pruning strategies to remove MPI library internals, line statements, and negligible code regions (less than 1\% of total runtime), our new \cpp{} \readapi{} ingested parallel profiles for the full 100,000-rank dataset in an average of 9.6878 seconds on Aurora.

This represents a paradigm shift in "time-to-insight". By evaluating pruning logic once on the summary profile and propagating it via binary search across all MPI ranks, \hpcanalysis{} efficiently imports critical code regions for massive-scale executions. Furthermore, our accelerated \cpp{} API eliminates the sampling dependencies that limited prior work. Table~\ref{tab:ingestion_performance} compares the original Python-based \readapi{} using \texttt{joblib.Parallel} and various task counts against our new \cpp{} API utilizing OpenMP multi-threading. On Aurora, each compute node contains 104 cores with 2 hardware threads per core; thus, in Table \ref{tab:ingestion_performance}, our \cpp{} API utilized 204 OpenMP threads to maximize hardware concurrency.

In Table~\ref{tab:ingestion_performance}, we measured performance when sampling up to 10,000 MPI ranks, as the Python API’s performance began to significantly degrade at that scale—requiring several minutes to process data that the \cpp{} API handled in approximately one second. The results indicate that the initial Python-based API suffered from poor scalability, whereas the \cpp{} implementation scaled with high efficiency. Ultimately, the \cpp{} API imported the entire 100,000-rank dataset in only 9.6878 seconds. All metrics in Table~\ref{tab:ingestion_performance} represent an average across 10 trials. While we performed the entire analysis on Aurora, users may also transfer collected measurements to systems with NVIDIA or AMD GPUs to leverage more mature GPU-accelerated backends.

\begin{table}[h]
\caption{Ingestion latency (s) for the 100,000-rank AMG execution on Aurora: Python (\texttt{joblib}) vs. accelerated \cpp{} (OpenMP) across sampling scales.}
\label{tab:ingestion_performance}
\centering
\resizebox{\columnwidth}{!}{%
\begin{tabular}{@{}lcccc@{}}
\toprule
\multirow{2}{*}{\textbf{Implementation}} & \multicolumn{4}{c}{\textbf{MPI Ranks Sampled}} \\ \cmidrule(l){2-5} 
 & \textbf{10} & \textbf{100} & \textbf{1,000} & \textbf{10,000} \\ \midrule
Python (16 tasks) & 3.7252 & 1.6838 & 18.3799 & 444.1241 \\
Python (32 tasks) & 0.2338 & 1.8088 & 16.2053 & 166.0623 \\
Python (64 tasks) & 0.2335 & 1.8954 & 15.7189 & 166.7260 \\
Python (128 tasks) & 0.2287 & 1.7087 & 16.2276 & 160.7161 \\ \midrule
\textbf{\cpp{} (204 threads)} & \textbf{0.1152} & \textbf{0.1224} & \textbf{0.2674} & \textbf{1.1265} \\ \bottomrule
\end{tabular}%
}
\end{table}

\begin{figure*}[!t]
\centering
\subfloat[\centering]{\includegraphics[width=.33\textwidth]{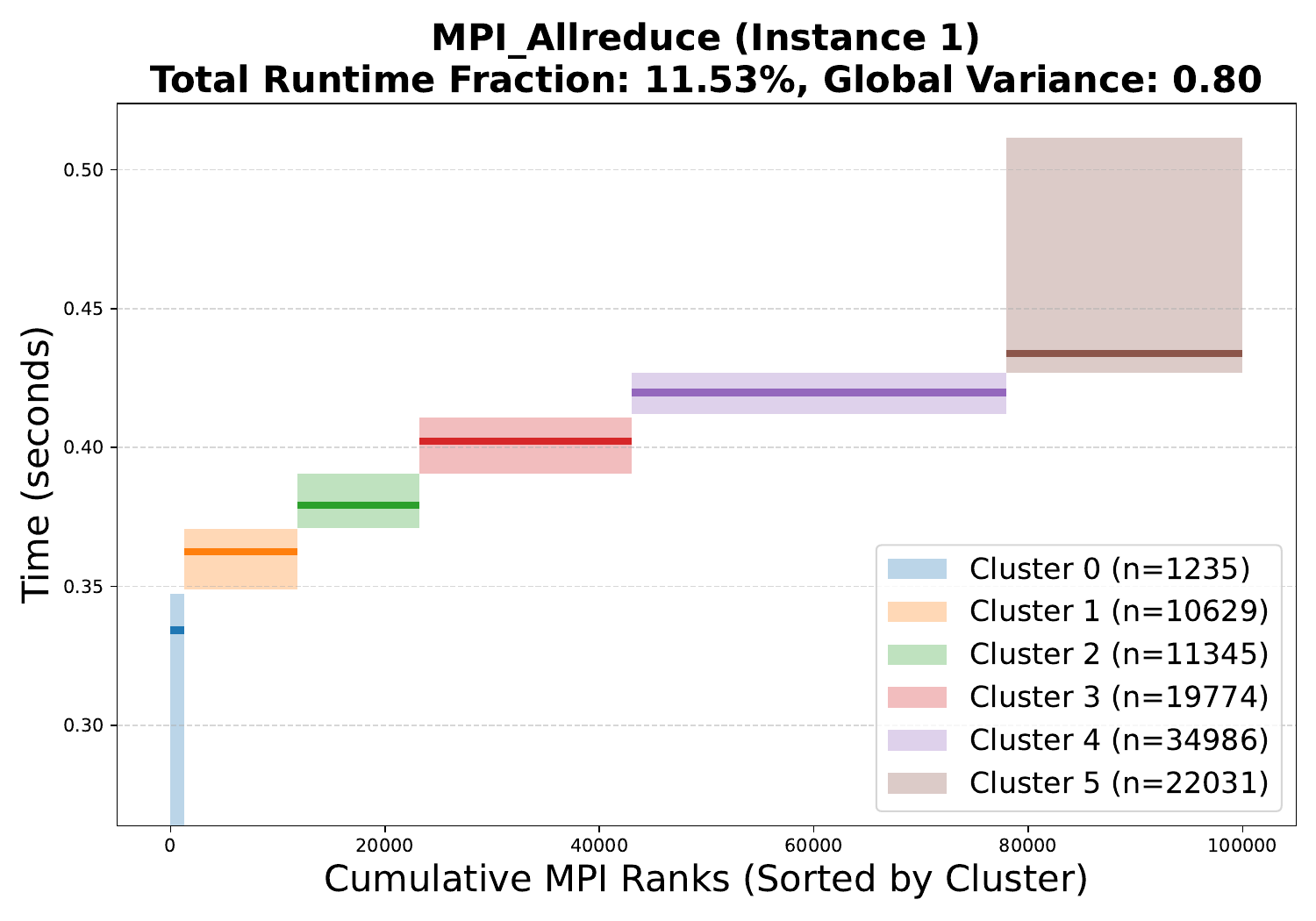}}
\subfloat[\centering]{\includegraphics[width=.33\textwidth]{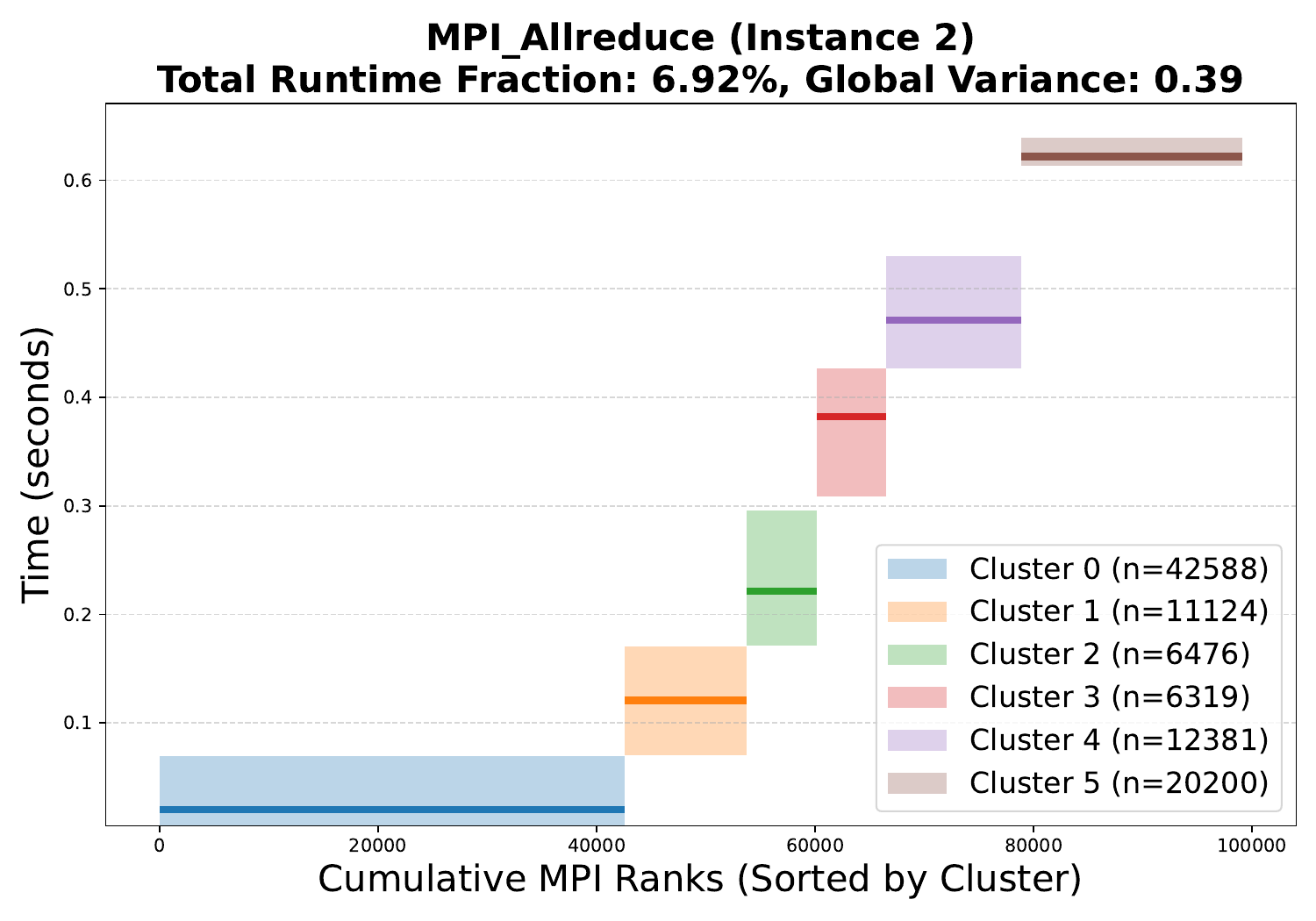}}
\subfloat[\centering]{\includegraphics[width=.33\textwidth]{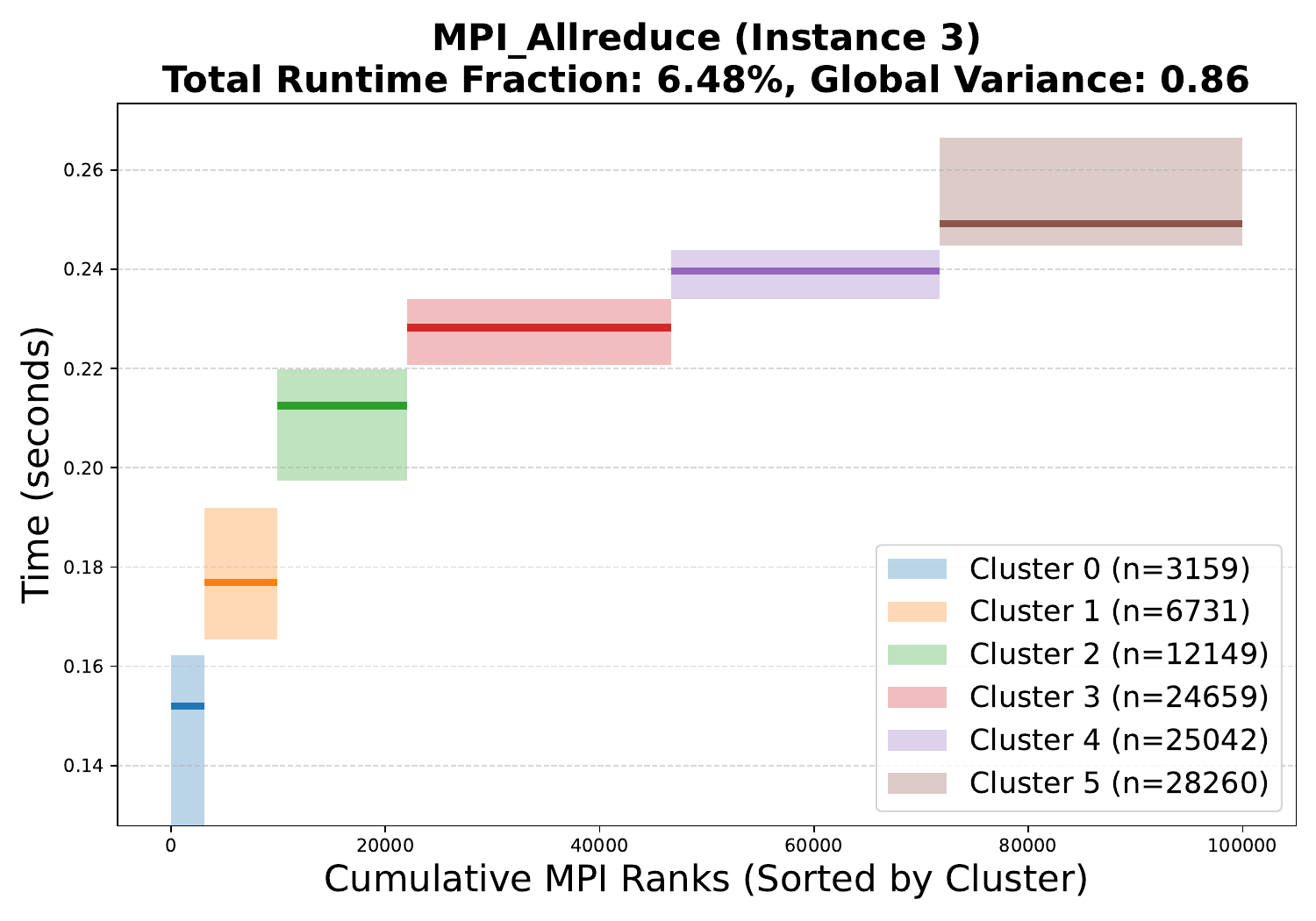}}\\
\subfloat[\centering]{\includegraphics[width=.33\textwidth]{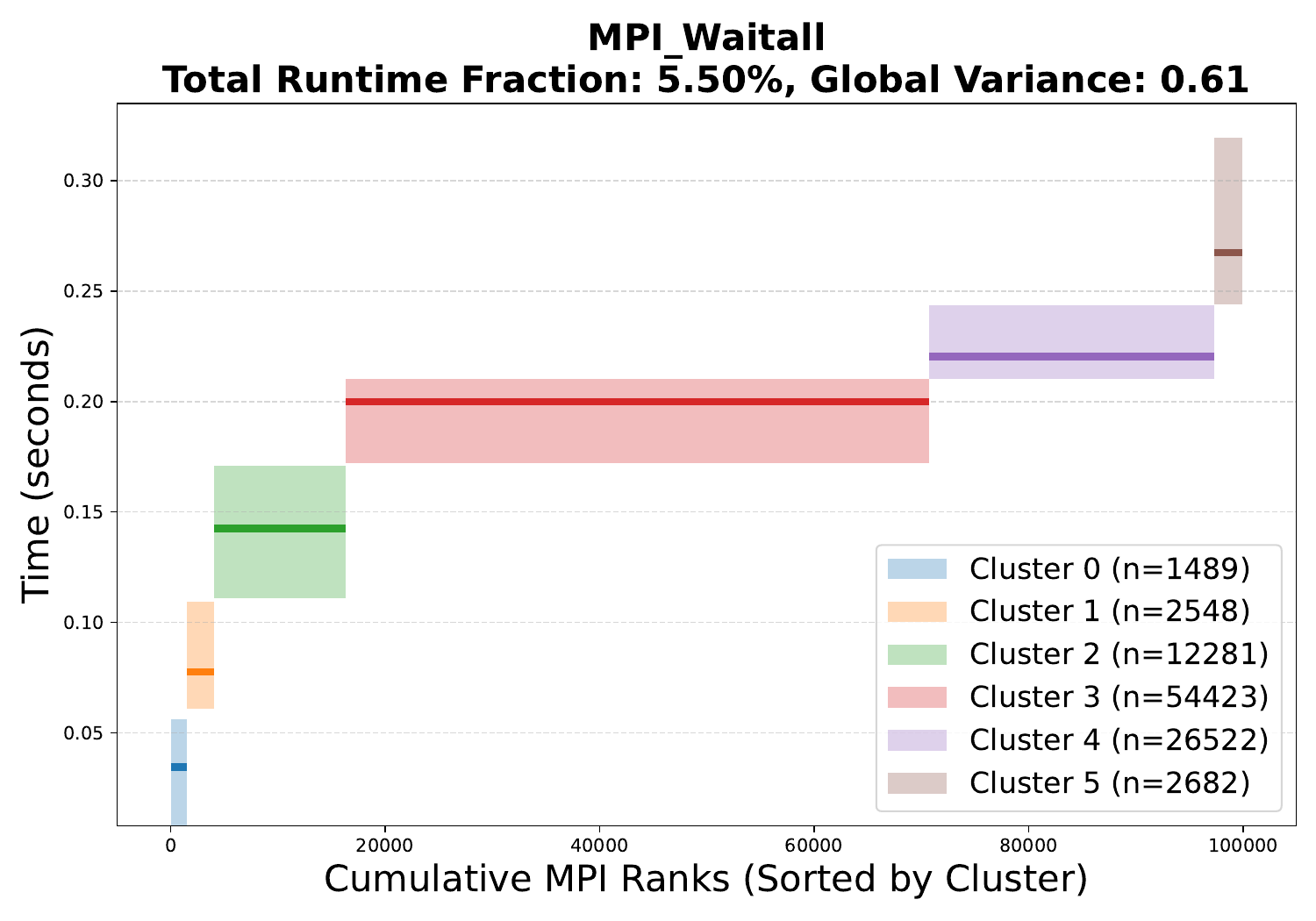}}
\subfloat[\centering]{\includegraphics[width=.33\textwidth]{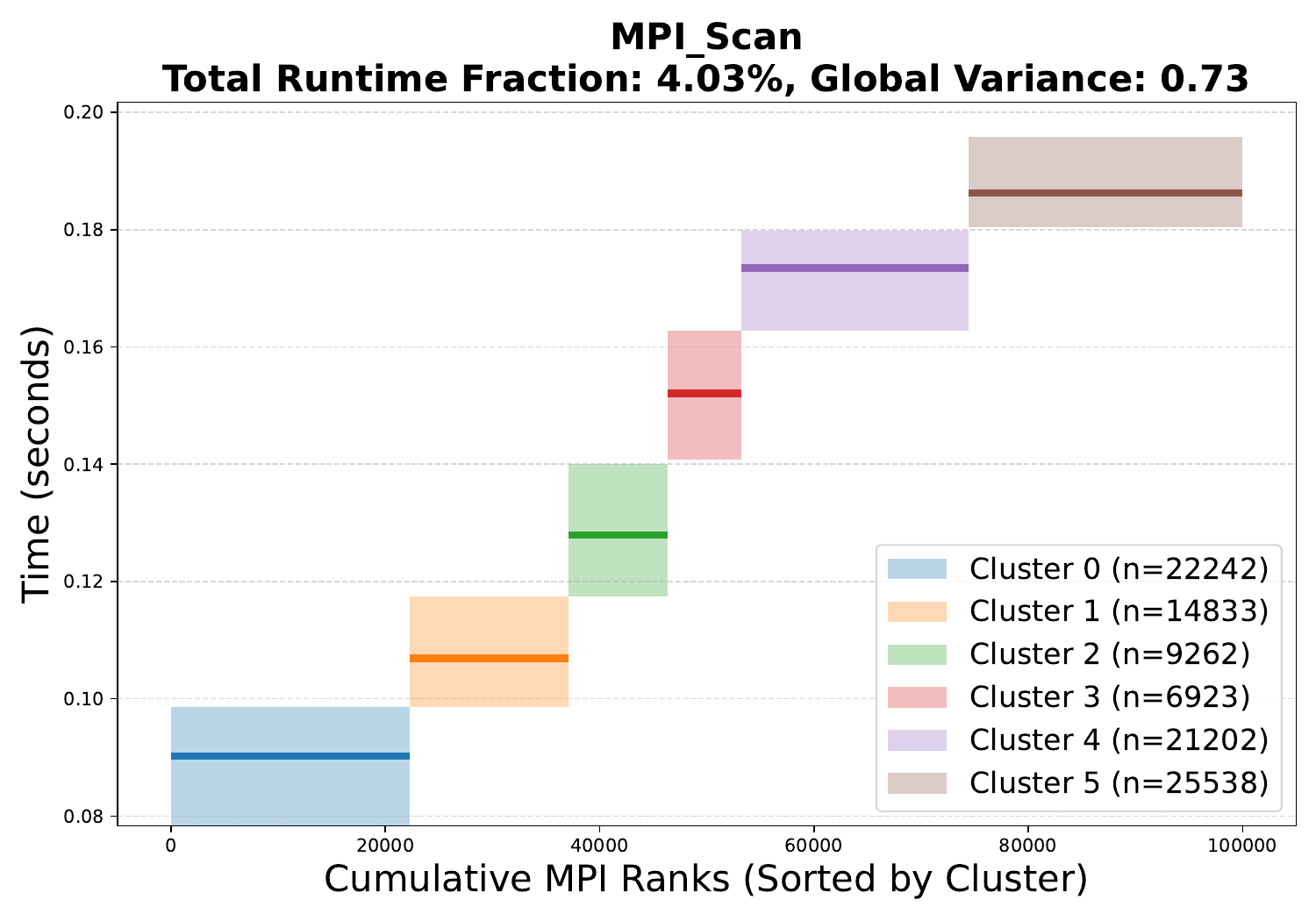}}
\subfloat[\centering]{\includegraphics[width=.33\textwidth]{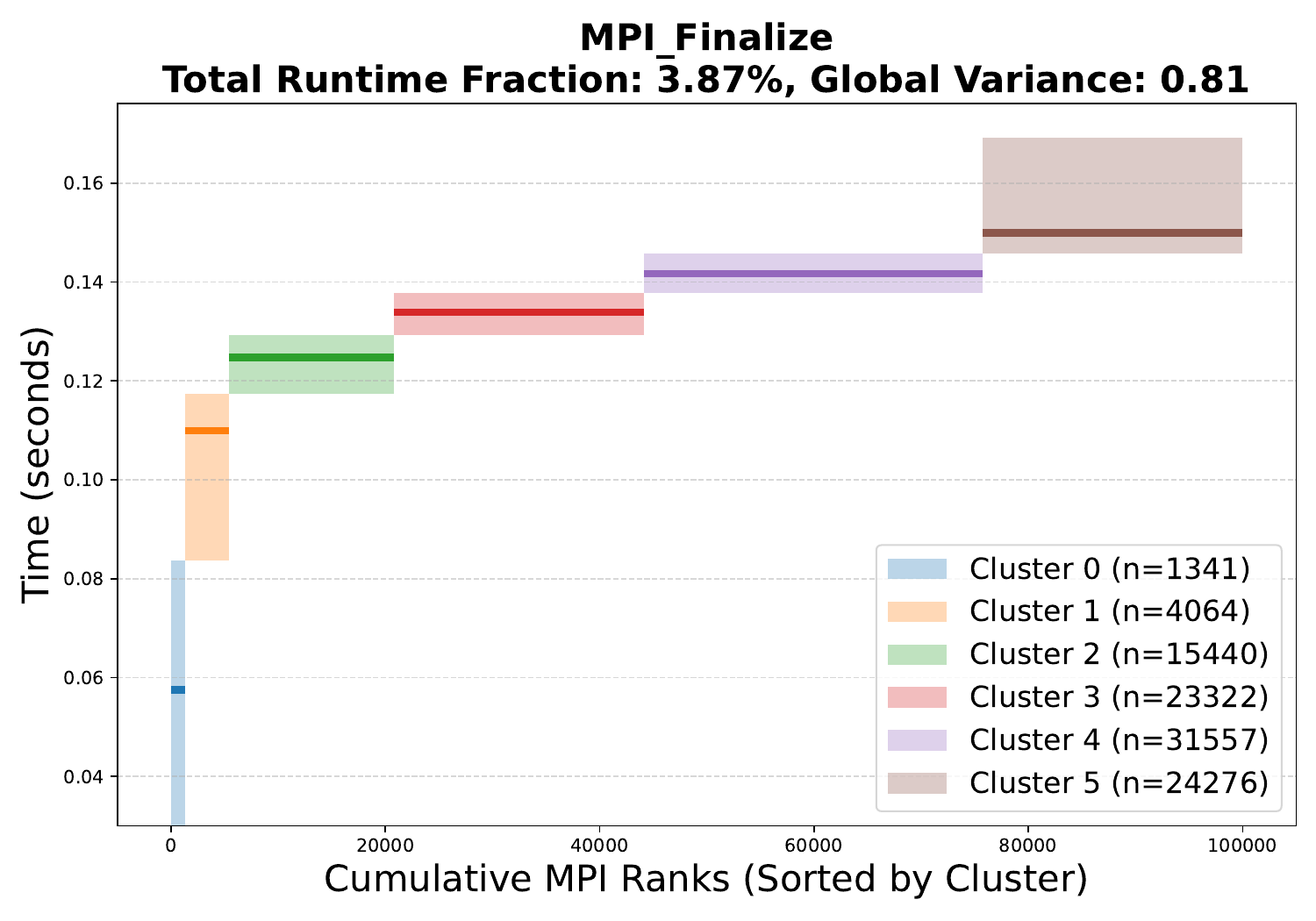}}
\caption{Performance distributions across ranks for the major MPI bottlenecks in the AMG exascale execution on Aurora.}
\label{fig:mpi}
\end{figure*}

\textbf{Quantifying Communication Imbalance.} Following the ingestion, we utilized the \queryapi{} to examine communication overhead. By querying \texttt{"summary"}, \texttt{"function(MPI\_*)"}, and \texttt{"cputime:sum (i)"}, we identified primary MPI bottlenecks within the summary profile. A subsequent request for rank-level granularity using \texttt{"rank"}, \texttt{"function(MPI\_*)"}, and \texttt{"cputime:prop (i)"} enabled us to calculate global balance ratios (average execution time across ranks divided by the maximum execution time across ranks) with only a few lines of Python code. Extracted performance data are represented as a DataFrame table where metric costs are attributed to calling context tree node IDs. Because a single MPI routine may be invoked from multiple locations, \hpcanalysis{} utilizes calling context tree node IDs to distinguish unique call sites. Fig.~\ref{fig:mpi} illustrates the performance distributions across ranks for the six primary MPI bottlenecks. While five routines were well-balanced (ratios near 1.0), the second instance of \texttt{MPI\_Allreduce} (Fig.~\ref{fig:mpi}b) exhibited a critical imbalance with a ratio of only 0.39. We clustered the 100,000 MPI ranks into six groups; for the critical \texttt{MPI\_Allreduce} instance, the brown cluster in Fig.~\ref{fig:mpi}b clearly identifies a significant performance outlier.

\textbf{Obtaining Calling Context.} To understand the source of the observed communication overhead, \hpcanalysis{} automatically reconstructs the full calling context for the calling context tree node ID. By maintaining a separate tree structure alongside the DataFrame, the framework performs highly efficient bottom-up traversal. Starting from the \texttt{MPI\_Allreduce} leaf ID, the framework identified the specific call chain: \texttt{hypre\_GMRESSetup} $\to$ \texttt{hypre\_BoomerAMGSetup} $\to$ \texttt{hypre\_ParCSRMatrixSetNumNonzeros\_core}.

\textbf{Node-Level Performance Correlation.} To investigate whether the observed communication overhead was tied to specific hardware, we mapped rank-level performance clusters to their compute nodes. By joining the DataFrame containing metric costs with the metadata DataFrame containing profile-level information (Fig.~\ref{fig:hpcanalysis}), we localized individual MPI ranks to logical compute nodes. When measuring applications across numerous compute nodes, HPCToolkit attributes nodes with their POSIX ID, enabling unique node-level attribution. For the critical \texttt{MPI\_Allreduce} instance, we first applied DBSCAN~\cite{dbscan} to the 100,000 MPI ranks, which partitioned the 1,000-node allocation into two distinct groups of 798 and 202 compute nodes with zero intersection. Interestingly, querying the total execution time across ranks (by replacing \texttt{"function(MPI\_*)"} with \texttt{"function(main)"}) yielded the exact same exclusive grouping: the first group averaged 3.12s, while the 202-node outlier group averaged 5.19s. We further tested cluster stability for the group of 798 compute nodes using K-Means~\cite{kmeans} across 2 to 5 sub-groups, finding minimal intersections:

\begin{itemize}
    \item \textbf{2 groups:} Only 2 compute nodes intersected.
    \item \textbf{3 groups:} 13 compute nodes intersected between groups 1 and 2; no others.
    \item \textbf{4 groups:} 13 compute nodes intersected between groups 1 and 2, 3 compute nodes between groups 2 and 3, and 13 compute nodes between groups 3 and 4.
    \item \textbf{5 groups:} Intersections were limited to 12 compute nodes between groups 2 and 3 and 16 compute nodes between groups 3 and 4.
\end{itemize}

These minimal intersections across a 1,000-node allocation suggest a high correlation between elevated MPI latency and specific logical compute nodes.

\begin{figure}
    \centering
    \includegraphics[width=.4\textwidth]{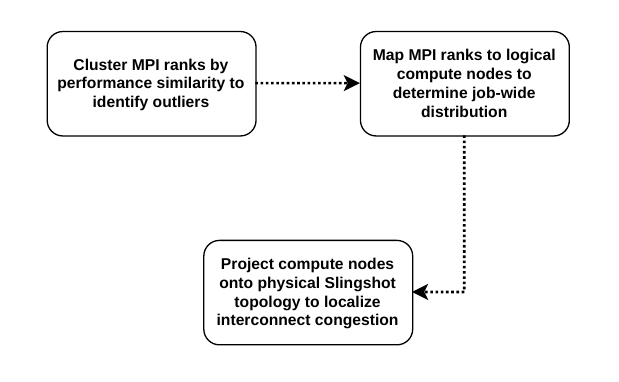}
    \caption{Topology-aware diagnostic workflow for localizing exascale interconnect congestion. The process maps logical performance outliers to physical interconnect coordinates.}
    \label{fig:congestion}
\end{figure}

\textbf{Interconnect Mapping and Congestion Analysis.} To localize the observed 202 outlier compute nodes, we implemented a workflow that maps collected POSIX IDs to physical Slingshot~\cite{slingshot} interconnect coordinates on Aurora. We first examined the MPI ranks to detect groups of similar behavior. We then utilized metadata DataFrame table to uniquely distinguish between compute nodes. While POSIX IDs enabled statistical grouping, they did not reveal the topographical location of the congestion. Therefore, we developed a workflow that maps POSIX IDs to Aurora's hexadecimal physical addresses, which follow formats such as:

\begin{lstlisting}[language=Python]
"x4109c0s0b0n0"
\end{lstlisting}

where coordinates represent racks ($x$), chassis ($c$), and slots ($s$). All strings end with $b0n0$, representing the single blade/node within the slot~\cite{aurora_slingshot}. By mapping the 202 outlier compute nodes to the interconnect, we determined that the congestion was not confined to a single rack or chassis. The affected nodes were distributed across 22 different racks; in some racks, entire chassis were affected, while in others, only a few showed degradation. By identifying physical coordinates of communication overhead at this scale, we prove that \hpcanalysis{} can efficiently examine exascale executions to detect, quantify, and physically locate network congestion within a production interconnect. Fig.~\ref{fig:congestion} illustrates the entire workflow we utilized to localize network congestion for the AMG exascale execution on Aurora.

\section{Conclusion}\label{sec:conclusion}
As high-performance computing enters the exascale era, the sheer volume and complexity of performance telemetry have outpaced the capabilities of traditional, CPU-bound analysis tools. In this paper, we presented an accelerated, heterogeneous infrastructure for the \hpcanalysis{} framework, leveraging a high-performance \cpp{} API and GPU parallelism to meet the throughput demands of exascale executions. With this enhanced foundation, we demonstrated the ability to ingest performance datasets for 100,000-rank executions in seconds and efficiently process them to detect critical performance inefficiencies. On the Aurora supercomputer, we integrated a topology-aware workflow that maps logical performance outliers to their physical locations within the Slingshot interconnect, localizing network congestion at the rack and chassis level during massive-scale executions.

Beyond raw throughput, we demonstrated how the programmatic interface of \hpcanalysis{} enables seamless integration with external tools to provide sophisticated analytical models. By integrating \hpcanalysis{} with the Thicket framework, we introduced a novel tri-dimensional performance model ($Node \times Trace \times Iteration$) that "re-materializes" iterative behavior within and across execution traces. This model enables the detection of iterative imbalances, information typically lost in aggregate performance profiles. With this model, we were able to perform fine-grained analysis of GPU idleness in production workloads and estimate potential performance gains on Frontier. Collectively, our contributions provide a scalable and extensible substrate for performance diagnostics at the absolute limit of modern supercomputing. We believe that the \hpcanalysis{} framework will have a transformative impact on the HPC community, driving critical advancements in the analysis and optimization of exascale computing systems. \hpcanalysis{} is an open-source project available at~\cite{hpcanalysis_code}.

\section{Acknowledgements}
This research was supported in part by UT-Battelle, LLC subcontract CW54422 (DOE Prime DE-AC05-00OR22275), LLNL subcontracts B665301 and B670681 (DOE Prime DE-AC52-07NA27344), ANL subcontract 4F-60094 (DOE Prime DE-AC02-06CHII357), and a contract from TotalEnergies E\&P Research \& Technology USA, LLC.

This research used resources of the Oak Ridge Leadership Computing Facility, a DOE Office of Science User Facility at Oak Ridge National Laboratory, supported under Contract DE-AC05-00OR22725. This research also used resources of the Argonne Leadership Computing Facility, a DOE Office of Science User Facility at Argonne National Laboratory, based on research supported by the U.S. DOE Office of Science–Advanced Scientific Computing Research Program, supported under Contract DE-AC02-06CH11357.

\bibliographystyle{IEEEtran}
\balance
\bibliography{references}

@misc{frontier_1, 
    title={{Frontier}},
    author={{Oak Ridge Leadership Computing Facilty}},
    url={https://www.olcf.ornl.gov/frontier},
    year={2026},
    note={{Accessed: April 2026}},
}

@misc{frontier_2,
    title={{Frontier User Guide}},
    author={{Oak Ridge Leadership Computing Facilty}},
    url={https://docs.olcf.ornl.gov/systems/frontier_user_guide.html},
    year={2026},
    note={{Accessed: April 2026}},
}

@misc{aurora_1, 
    title={{Aurora}},
    author={{Argonne Leadership Computing Facility}},
    url={https://www.alcf.anl.gov/aurora},
    year={2026},
    note={{Accessed: April 2026}},
}

@misc{aurora_2,
    title={{Aurora User Guide}},
    author={{Argonne Leadership Computing Facility}},
    url={https://docs.alcf.anl.gov/aurora/getting-started-on-aurora},
    year={2026},
    note={{Accessed: April 2026}},
}

@misc{elcapitan,
    title={{El Capitan: Preparing for NNSA’s first exascale machine}},
    author={{Lawrence Livermore National Laboratory}},
    url={https://asc.llnl.gov/exascale/el-capitan},
    year={2026},
    note={{Accessed: April 2026}},
}

@article{hpctoolkit_cpu,
    title={{HPCToolkit: Tools for performance analysis of optimized parallel programs}},
    author={Adhianto, Laksono and Banerjee, Sinchan and Fagan, Mike and Krentel, Mark and Marin, Gabriel and Mellor-Crummey, John and Tallent, Nathan R},
    journal={Concurrency and Computation: Practice and Experience},
    volume={22},
    number={6},
    pages={685--701},
    year={2010},
    publisher={Wiley Online Library},
    url={https://www.doi.org/10.1002/cpe.1553},
}

@article{hpctoolkit_gpu,
    title={{Measurement and analysis of GPU-accelerated applications with HPCToolkit}},
    author={Zhou, Keren and Adhianto, Laksono and Anderson, Jonathon and Cherian, Aaron and Grubisic, Dejan and Krentel, Mark and Liu, Yumeng and Meng, Xiaozhu and Mellor-Crummey, John},
    journal={Parallel Computing},
    volume={108},
    pages={102837},
    year={2021},
    publisher={Elsevier},
    url={https://www.doi.org/10.1016/j.parco.2021.102837},
}

@article{hpctoolkit_refinement,
    title={{Refining HPCToolkit for application performance analysis at exascale}},
    author={Adhianto, Laksono and Anderson, Jonathon and Barnett, Robert Matthew and Grbic, Dragana and Indic, Vladimir and Krentel, Mark and Liu, Yumeng and Milakovi{\'c}, Sr{\dj}an and Phan, Wileam and Mellor-Crummey, John},
    journal={The International Journal of High Performance Computing Applications},
    volume={38},
    number={6},
    pages={612--632},
    year={2024},
    publisher={Sage Publications Sage UK: London, England},
    url={https://www.doi.org/10.1177/10943420241277839},
}

@inproceedings{hpcanalysis,
    title={{Analyzing the Performance of Applications at Exascale}},
    author={Grbic, Dragana and Mellor-Crummey, John},
    booktitle={Proceedings of the 39th ACM International Conference on Supercomputing},
    pages={792--806},
    year={2025},
    url={https://www.doi.org/10.1145/3721145.3730417},
}

@misc{nsight_systems, 
    title={{NVIDIA Nsight Systems}},
    author={{NVIDIA Corporation}},
    url={https://developer.nvidia.com/nsight-systems},
    year={2026},
    note={{Accessed: April 2026}},
}

@misc{nsight_compute, 
    title={{NVIDIA Nsight Compute}},
    author={{NVIDIA Corporation}},
    url={https://developer.nvidia.com/nsight-compute},
    year={2026},
    note={{Accessed: April 2026}},
}

@inproceedings{tau,
    title={{Advances in the TAU performance system}},
    author={Malony, Allen and Shende, Sameer and Spear, Wyatt and Lee, Chee Wai and Biersdorff, Scott},
    booktitle={Tools for High Performance Computing 2011: Proceedings of the 5th International Workshop on Parallel Tools for High Performance Computing, September 2011, ZIH, Dresden},
    pages={119--130},
    year={2012},
    organization={Springer},
    url={https://www.doi.org/10.1007/978-3-642-31476-6_10},
}

@inproceedings{score_p,
    title={{Score-P: A unified performance measurement system for petascale applications}},
    author={Mey, Dieter An and Biersdorf, Scott and Bischof, Christian and Diethelm, Kai and Eschweiler, Dominic and Gerndt, Michael and Kn{\"u}pfer, Andreas and Lorenz, Daniel and Malony, Allen and Nagel, Wolfgang E and others},
    booktitle={Competence in High Performance Computing 2010: Proceedings of an International Conference on Competence in High Performance Computing, June 2010, Schloss Schwetzingen, Germany},
    pages={85--97},
    year={2011},
    organization={Springer},
    url={https://www.doi.org/10.1007/978-3-642-24025-6_8},
}

@inproceedings{caliper,
    title={{Caliper: performance introspection for HPC software stacks}},
    author={Boehme, David and Gamblin, Todd and Beckingsale, David and Bremer, Peer-Timo and Gimenez, Alfredo and LeGendre, Matthew and Pearce, Olga and Schulz, Martin},
    booktitle={SC'16: Proceedings of the International Conference for High Performance Computing, Networking, Storage and Analysis},
    pages={550--560},
    year={2016},
    organization={IEEE},
    url={https://www.doi.org/10.1109/SC.2016.46},
}

@inproceedings{extrae,
    title={{Experiences on the characterization of parallel applications in embedded systems with extrae/paraver}},
    author={Munera, Adrian and Royuela, Sara and Llort, Germ{\'a}n and Mercadal, Estanislao and Wartel, Franck and Qui{\~n}ones, Eduardo},
    booktitle={Proceedings of the 49th International Conference on Parallel Processing},
    pages={1--11},
    year={2020},
    url={https://www.doi.org/10.1145/3404397.3404440},
}

@article{scalasca,
    title={{The Scalasca performance toolset architecture}},
    author={Geimer, Markus and Wolf, Felix and Wylie, Brian JN and {\'A}brah{\'a}m, Erika and Becker, Daniel and Mohr, Bernd},
    journal={Concurrency and computation: Practice and experience},
    volume={22},
    number={6},
    pages={702--719},
    year={2010},
    publisher={Wiley Online Library},
    url={https://www.doi.org/10.1002/cpe.1556},
}

@inproceedings{vampir,
    title={{Parallel performance engineering using Score-P and Vampir}},
    author={Williams, William and Brunst, Holger},
    booktitle={Companion of the 2023 ACM/SPEC International Conference on Performance Engineering},
    pages={121--125},
    year={2023},
    url={https://www.doi.org/10.1145/3578245.3583715},
}

@article{paraver,
    title={{15+ years of joint parallel application performance analysis/tools training with Scalasca/Score-P and Paraver/Extrae toolsets}},
    author={Wylie, Brian JN and Gim{\'e}nez, Judit and Feld, Christian and Geimer, Markus and Llort, Germ{\'a}n and Mendez, Sandra and Mercadal, Estanislao and Visser, Anke and Garc{\'\i}a-Gasulla, Marta},
    journal={Future Generation Computer Systems},
    volume={162},
    pages={107472},
    year={2025},
    publisher={Elsevier},
    url={https://www.doi.org/10.1016/j.future.2024.07.050},
}

@inproceedings{hatchet,
    title={{Hatchet: Pruning the overgrowth in parallel profiles}},
    author={Bhatele, Abhinav and Brink, Stephanie and Gamblin, Todd},
    booktitle={Proceedings of the International Conference for High Performance Computing, Networking, Storage and Analysis},
    pages={1--21},
    year={2019},
    url={https://www.doi.org/10.1145/3295500.3356219},
}

@incollection{pandas,
    title={{Python data analysis with pandas}},
    author={Bernard, Joey},
    booktitle={Python recipes handbook: A problem-solution approach},
    pages={37--48},
    year={2016},
    publisher={Springer},
    url={https://www.doi.org/10.1007/978-1-4842-0241-8_5},
}

@inproceedings{thicket,
    title={{Thicket: seeing the performance experiment forest for the individual run trees}},
    author={Brink, Stephanie and McKinsey, Michael and Boehme, David and Scully-Allison, Connor and Lumsden, Ian and Hawkins, Daryl and Burgess, Treece and Lama, Vanessa and L{\"u}ttgau, Jakob and Isaacs, Katherine E and others},
    booktitle={Proceedings of the 32nd International Symposium on High-Performance Parallel and Distributed Computing},
    pages={281--293},
    year={2023},
    url={https://www.doi.org/10.1145/3588195.3592989},
}

@article{pipit,
    title={{Pipit: Scripting the analysis of parallel execution traces}},
    author={Bhatele, Abhinav and Dhakal, Rakrish and Movsesyan, Alexander and Ranjan, Aditya K and Cankur, Onur},
    journal={arXiv preprint arXiv:2306.11177},
    year={2023},
    url={https://www.doi.org/10.48550/arXiv.2306.11177},
}

@article{projections,
    title={{Scaling applications to massively parallel machines using projections performance analysis tool}},
    author={Kale, Laxmikant V and Zheng, Gengbin and Lee, Chee Wai and Kumar, Sameer},
    journal={Future Generation Computer Systems},
    volume={22},
    number={3},
    pages={347--358},
    year={2006},
    publisher={Elsevier},
    url={https://www.doi.org/10.1016/j.future.2004.11.020},
}

@inproceedings{open_trace,
    title={{Introducing the open trace format (OTF)}},
    author={Kn{\"u}pfer, Andreas and Brendel, Ronny and Brunst, Holger and Mix, Hartmut and Nagel, Wolfgang E},
    booktitle={International Conference on Computational Science},
    pages={526--533},
    year={2006},
    organization={Springer},
    url={https://www.doi.org/10.1007/11758525_71},
}

@inproceedings{hpctoolkit_ics,
    title={{Preparing for performance analysis at exascale}},
    author={Anderson, Jonathon and Liu, Yumeng and Mellor-Crummey, John},
    booktitle={Proceedings of the 36th ACM International Conference on Supercomputing},
    pages={1--13},
    year={2022},
    url={https://www.doi.org/10.1145/3524059.3532397},
}

@misc{jupyter,
    title={{Project Jupyter: Interactive Computing across Programming Languages}},
    author={{Project Jupyter}},
    url={https://jupyter.org},
    year={2026},
    note={{Accessed: April 2026}},
}

@article{lammps,
    title={{LAMMPS-a flexible simulation tool for particle-based materials modeling at the atomic, meso, and continuum scales}},
    author={Thompson, Aidan P and Aktulga, H Metin and Berger, Richard and Bolintineanu, Dan S and Brown, W Michael and Crozier, Paul S and In't Veld, Pieter J and Kohlmeyer, Axel and Moore, Stan G and Nguyen, Trung Dac and others},
    journal={Computer physics communications},
    volume={271},
    pages={108171},
    year={2022},
    publisher={Elsevier},
    url={https://www.doi.org/10.1016/j.cpc.2021.108171},
}

@misc{mmap,
    title={{mmap(2) — Linux Manual Page}},
    author={{Linux Man-Pages Project}},
    url={https://man7.org/linux/man-pages/man2/mmap.2.html},
    year={2026},
    note={{Accessed: April 2026}},
}

@misc{joblib,
    title={{Joblib: Running Python Functions as Pipeline Jobs}},
    author={{Joblib Development Team}},
    url={https://joblib.readthedocs.io},
    year={2026},
    note={{Accessed: April 2026}},
}

@misc{pybind11,
    title={{pybind11: Seamless Operability Between C++11 and Python}},
    author={{pybind11 Development Team}},
    url={https://pybind11.readthedocs.io},
    year={2026},
    note={{Accessed: April 2026}},
}

@article{numpy,
    title={{Array programming with NumPy}},
    author={Harris, Charles R and Millman, K Jarrod and Van Der Walt, St{\'e}fan J and Gommers, Ralf and Virtanen, Pauli and Cournapeau, David and Wieser, Eric and Taylor, Julian and Berg, Sebastian and Smith, Nathaniel J and others},
    journal={nature},
    volume={585},
    number={7825},
    pages={357--362},
    year={2020},
    publisher={Nature Publishing Group UK London},
    url={https://www.doi.org/10.1038/s41586-020-2649-2},
}

@misc{cudf,
    title={{cuDF: GPU-Accelerated Pandas-like DataFrames}},
    author={{RAPIDS Development Team}},
    url={https://docs.rapids.ai/api/cudf/stable},
    year={2026},
    note={{Accessed: April 2026}},
}

@misc{hipdf,
    title={{hipDF: HIP-based DataFrames for AMD GPUs}},
    author={{AMD ROCm Development Team}},
    url={https://rocm.docs.amd.com/projects/hipDF/en/latest},
    year={2026},
    note={{Accessed: April 2026}},
}

@misc{dpnp,
    title={{dpnp: NumPy-compliant Interface for Data Parallel C++ (DPC++)}},
    author={{Intel oneAPI Development Team}},
    url={https://intelpython.github.io/dpnp},
    year={2026},
    note={{Accessed: April 2026}},
}

@misc{amg,
    title={{AMG: Algebraic Multi-Grid Parallel Iterative Solver Benchmark}},
    author={{Lawrence Livermore National Laboratory}},
    url={https://www.osti.gov/biblio/1389816},
    year={2026},
    note={{Accessed: April 2026}},
}

@misc{polaris_1, 
    title={{Polaris}},
    author={{Argonne Leadership Computing Facility}},
    url={https://www.alcf.anl.gov/polaris},
    year={2026},
    note={{Accessed: April 2026}},
}

@misc{polaris_2,
    title={{Polaris User Guide}},
    author={{Argonne Leadership Computing Facility}},
    url={https://docs.alcf.anl.gov/polaris/getting-started},
    year={2026},
    note={{Accessed: April 2026}},
}

@misc{cuda,
    title={{NVIDIA CUDA Toolkit}},
    author={{NVIDIA Corporation}},
    url={https://developer.nvidia.com/cuda-toolkit},
    year={2026},
    note={Accessed: April 2026},
}

@misc{rocm,
    title={{ROCm: Open Software Platform for GPU Compute}},
    author={{Advanced Micro Devices, Inc.}},
    url={https://rocm.docs.amd.com},
    year={2026},
    note={Accessed: April 2026},
}

@misc{oneapi,
    title={{Intel oneAPI Toolkits}},
    author={{Intel Corporation}},
    url={https://www.intel.com/content/www/us/en/developer/tools/oneapi/overview.html},
    year={2026},
    note={Accessed: April 2026},
}

@misc{levelzero,
    title={{Intel oneAPI Level Zero Specification}},
    author={{Intel Corporation}},
    url={https://oneapi-src.github.io/level-zero-spec},
    year={2026},
    note={Accessed: April 2026},
}

@article{gamess,
    title={{Recent developments in the general atomic and molecular electronic structure system}},
    author={Barca, Giuseppe MJ and Bertoni, Colleen and Carrington, Laura and Datta, Dipayan and De Silva, Nuwan and Deustua, J Emiliano and Fedorov, Dmitri G and Gour, Jeffrey R and Gunina, Anastasia O and Guidez, Emilie and others},
    journal={The Journal of chemical physics},
    volume={152},
    number={15},
    year={2020},
    publisher={AIP Publishing},
    url={https://www.doi.org/10.1063/5.0005188},
}

@inproceedings{dbscan,
    title={{DBSCAN clustering algorithm based on density}},
    author={Deng, Dingsheng},
    booktitle={2020 7th international forum on electrical engineering and automation (IFEEA)},
    pages={949--953},
    year={2020},
    organization={IEEE},
    url={https://www.doi.org/10.1109/IFEEA51475.2020.00199},
}

@article{kmeans,
    title={{Unsupervised K-means clustering algorithm}},
    author={Sinaga, Kristina P and Yang, Miin-Shen},
    journal={IEEE access},
    volume={8},
    pages={80716--80727},
    year={2020},
    publisher={IEEE},
    url={https://www.doi.org/10.1109/ACCESS.2020.2988796},
}

@inproceedings{slingshot,
    title={{An in-depth analysis of the slingshot interconnect}},
    author={De Sensi, Daniele and Di Girolamo, Salvatore and McMahon, Kim H and Roweth, Duncan and Hoefler, Torsten},
    booktitle={SC20: International Conference for High Performance Computing, Networking, Storage and Analysis},
    pages={1--14},
    year={2020},
    organization={IEEE},
    url={https://www.doi.org/10.1109/SC41405.2020.00039},
}

@misc{aurora_slingshot,
    title={{Running Jobs on Aurora}},
    author={{Argonne Leadership Computing Facility}},
    url={https://docs.alcf.anl.gov/aurora/running-jobs-aurora},
    year={2026},
    note={{Accessed: April 2026}},
}

@misc{hpcanalysis_code, 
    title={{hpcanalysis}},
    author={The HPCToolkit Project},
    howpublished={\url{https://gitlab.com/draganaurosgrbic/hpcanalysis}},
    year={2026},
}

\end{document}